\documentclass[prd, twocolumn, nofootinbib, superscriptaddress]{revtex4-2}

\usepackage{aas_macros}
\usepackage{graphicx}
\usepackage[caption=false]{subfig}
\usepackage{siunitx}
\usepackage{mathrsfs}
\usepackage{amsmath,amssymb}
\usepackage{bm}
\usepackage{braket}
\usepackage{listings}
\usepackage{cases}
\usepackage{here}
\usepackage{comment}
\usepackage{soul}
\usepackage{cancel}
\usepackage{cases}
\usepackage[utf8]{inputenc}
\usepackage{url}
\usepackage{longtable}
\usepackage[normalem]{ulem}
\usepackage{xspace}
\usepackage[colorlinks=true
,urlcolor=blue
,anchorcolor=blue
,citecolor=blue
,filecolor=blue
,linkcolor=blue
,menucolor=blue
,linktocpage=true
,pdfproducer=medialab
,pdfa=true
]{hyperref}
\usepackage{animate}

\begin{document}


\title{Modeling the core-halo mass relation in fuzzy dark matter halos}

\author{Hiroki Kawai}
\email{hiroki.kawai@phys.s.u-tokyo.ac.jp}
\affiliation{Department of Physics, School of Science, The University of Tokyo, Bunkyo, Tokyo 113-0033, Japan}
\affiliation{Center for Frontier Science, Chiba University, 1-33 Yayoicho, Inage, Chiba 263-8522, Japan}
\affiliation{INAF – Osservatorio Astronomico di Bologna, via Ranzani 1, 40127 Bologna, Italy}

\author{Ayuki Kamada}
\affiliation{Institute of Theoretical Physics, Faculty of Physics, University of Warsaw, ul. Pasteura 5, PL–02–093 Warsaw, Poland}
\affiliation{Kavli Institute for the Physics and Mathematics of the Universe (Kavli IPMU, WPI), The University of Tokyo, Chiba 277-8583, Japan}
\affiliation{Center for Theoretical Physics of the Universe, Institute for Basic Science (IBS), Daejeon 34126, Korea}

\author{Kohei Kamada}
\affiliation{School of Fundamental Physics and Mathematical Sciences, Hangzhou Institute for Advanced Study, University of Chinese Academy of Sciences (HIAS-UCAS), Hangzhou 310024, China}
\affiliation{International Centre for Theoretical Physics Asia-Pacific (ICTP-AP), Beijing/Hangzhou, China}
\affiliation{Research Center for the Early Universe (RESCEU), Graduate School of Science, The University of Tokyo, Hongo 7-3-1 Bunkyo-ku, Tokyo 113-0033, Japan}

\author{Naoki Yoshida}
\affiliation{Department of Physics, School of Science, The University of Tokyo, Bunkyo, Tokyo 113-0033, Japan}
\affiliation{Kavli Institute for the Physics and Mathematics of the Universe (Kavli IPMU, WPI), The University of Tokyo, Chiba 277-8583, Japan}
\affiliation{Institute for Physics of Intelligence, School of Science, The University of Tokyo, Bunkyo, Tokyo 113-0033, Japan}

\date{\today}

\begin{abstract}
Fuzzy dark matter (FDM) is an intriguing candidate alternative to the standard cold dark matter (CDM).
The FDM model predicts that dark halos have characteristic core structures generated by the effect of quantum pressure, which is different from the structure of CDM halos.
We devise a semi-analytic model of a FDM halo density profile by assuming that the density distribution results from the redistribution of mass in a halo with the Navarro-Frenk-White profile. 
We calculate the mass redistribution radius by considering dynamical relaxation within the FDM halo.
We adopt a concentration-halo mass relation with lower concentration compared to that in the CDM model below the half mode mass, which originates from the suppressed matter density fluctuations at small length scales. 
Our model reproduces the core-halo mass relation (CHMR) found in the numerical simulation of \citet{2014NatPh..10..496S} at $z<1$. 
We show that the CHMR is well described by a double power law, unlike previous studies that approximate it by a single power law.
Our model predictions are in reasonable agreement with the results of the largest FDM simulation of \citet{2021MNRAS.506.2603M} at $z=3$.
We find that the core mass for a given halo mass follows the log-normal distribution, both in our model and in the simulation results for the first time, and quantitatively compare the variance of the distribution among them.
Although our model does not fully explain the scatter of the CHMR, we show the scatter of the concentration-halo mass relation sizably contributes to them.
\end{abstract}

\keywords{Dark Matter}

\maketitle

\section{Introduction}
Dark matter is one of the major components of our universe and plays an important role in structure formation.
While the nature of dark matter is still unknown, it is generally thought that it is cold, and interacts with itself and with other matter only through gravity.
The so-called $\Lambda$ Cold Dark Matter ($\Lambda$CDM) model explains a broad range of observations ({\it{e.g.,}} \citet{2004ApJ...606..702T, 2019MNRAS.489.2247C}), but the success is limited to structures at large length scales with $k < 1\  {\rm Mpc}^{-1}$, where $k$ is the wavenumber of density fluctuations. 
There remain several discrepancies between the prediction of the standard model and observations at smaller length scales, known as small-scale problems (\citet{2017ARA&A..55..343B} for a recent review).
Often discussed are the core-cusp problem \citep{1994Natur.370..629M}, the missing satellite problem \citep{1999ApJ...524L..19M}, the diversity problem of rotation curves \citep{2015MNRAS.452.3650O}, and the too-big-to-fail problem \citep{2011MNRAS.415L..40B}.

Fuzzy dark matter (FDM) has been invoked as a promising alternative to CDM to alleviate the small-scale problems.
FDM is thought to consist of scalar particles minimally coupled to gravity with negligible self-interaction whose mass is around $mc^{2} \simeq 10^{-23}-10^{-21}\ {\rm eV}$, where $c$ is the speed of light. 
Elementary particles with such a small mass have a large de Broglie wavelength of $\mathcal{O}(1)\ \rm kpc$, and thus their wave nature significantly affects structure at $k > 1\ \rm kpc^{-1}$.
Since the de Broglie wavelength is larger than the typical separation between dark matter particles in dense regions, FDM exists in the form of a coherent state.
Recently, cosmological simulations of structure formation in FDM models have been performed. Two unique features of the FDM halos have been identified; there exist soliton cores and granular structures ({\it{e.g.,}} \citet{2014NatPh..10..496S, 2018PhRvD..97h3519M,2021MNRAS.506.2603M}).
The central soliton core in the FDM halo yields a cored density profile, which is surrounded by the Navarro-Freck-White (NFW) profile. 
The density profile of the soliton core can be expressed by that of the ground state solution of the Schr\"{o}dinger-Poisson (SP) equation, while the outer region of the FDM halo may be considered as a superposition of other energy eigenstates ({\it i.e.}, excited states).
\citet{2022PhRvD.105b3512Y} and \citet{2022PhRvD.105j3506Z} characterize the FDM halo by decomposing it into multiple eigenstates and following the time evolution of the contribution of each state.
To be precise, the eigenstates are not exactly the same as those obtained from the original SP equation because the time evolution of the potential is not considered. Instead, it is replaced with the time-averaged potential in \citet{2022PhRvD.105b3512Y} and \citet{2022PhRvD.105j3506Z}.
Although the multiple states can co-exist in the central region, it is found that the ground state is dominant near the center, and thus the density distribution of the central region is accurately described by that of the ground state solution.
Often found granular structures are generated by the interference of FDM, and exist all over within and around halos.
The characteristic size of the granular structures corresponds to the de Broglie wavelength.
Understanding these unique features is important to study FDM when we compare them with observational data.

The FDM halos are characterized by the so-called core-halo mass relation (CHMR), that is, the core mass as a function of the halo mass.
While many numerical simulations predict that the soliton core exists in the center of the FDM halos, the CHMR obtained so far is not consistent among the literature and thus is not well understood.
For instance, single powers of $M_{\rm c} \propto M_{\rm h}^{\alpha}$ are obtained with $\alpha = 1/3$ \citep{2014NatPh..10..496S}, $\alpha = 5/9$ \citep{2018PhRvD..97h3519M}, and $\alpha = 9/10$ \citep{2022MNRAS.511..943C}.
\citet{2022MNRAS.511..943C} argues that there exists scatter in the CHMR which may be the result of different degrees of tidal disruption on the FDM halos originating from the different simulation box sizes; halos in smaller boxes tend to be more tidally disrupted.
However, the largest cosmological simulation of \citet{2021MNRAS.506.2603M}, which contains many FDM halos, also shows a substantial scatter in the CHMR.
It might indicate that the scatter in the CHMR exists intrinsically regardless of the artificial factor.

It is expensive to perform the FDM simulations because of the necessity of high resolution in both space and time and therefore (semi-)analytic approaches are needed to understand the CHMR. 
\citet{2014PhRvL.113z1302S} considers that the physical size of the soliton core can be related to the velocity dispersion of the host halo through the uncertainty principle. 
The single-power law relation of $M_{\rm c} \propto M_{\rm h}^{1/3}$ is found, which is consistent with the simulation result of \citet{2014NatPh..10..496S}.
This relation can be understood as the total energy per mass is the same in the soliton core and in the halo \citep{2018PhRvD..98h3027B}.
\citet{2022PhRvD.106j3532T} derives the CHMR by considering the NFW profile as a background (but ignoring the self-gravity of the soliton core), and it seems that the scatter in the concentration-halo mass relation in the NFW profile contributes to a part (but not whole) scatter of the CHMR.
Despite these detailed studies, there are still no (semi-)analytic models that can fully explain the scatter seen in the simulation results. 

In the present paper, we develop a semi-analytic model of the CHMR in the FDM model. 
Our model assumes that the density distribution of FDM results from redistributing the mass in the cusped CDM halo in the central region where the wave nature of FDM is important. 
We first use the concentration-halo mass relation for the FDM model, which is modified from that of the CDM model below the half-mode mass \citep{2016ApJ...818...89S} owing to suppression of the linear power spectrum \citep{2022MNRAS.515.5646D, 2022MNRAS.517.1867L}, and construct the NFW profile for a given halo mass. 
We then adopt a relaxation time condition (and also a hydrostatic equilibrium condition) to calculate a characteristic radius, below which the dynamical relaxation occurs within the halo age to form the soliton core.
After determining the characteristic radius, we construct the soliton core by imposing a mass-matching condition around the matching radius which is defined as $\mathcal{O}$(1) times the characteristic radius.
We determine the ratio of the characteristic radius and the matching radius so that it can recover the CHMR obtained from the simulation of \citet{2014NatPh..10..496S} at redshift $z<1$, and that obtained from \citet{2021MNRAS.506.2603M} at redshift $z=3$, respectively.
In our model, the ratio of the characteristic radius and matching radius is assumed to be constant for all the FDM halos at a given redshift, but we allow redshift dependence.
We find that the CHMR can be expressed roughly by a double power law as a function of halo mass, where the turnover mass corresponds to the half-mode halo mass.
We study the parameter dependence of the CHMR, including the scatter of the concentration-halo mass relation, the FDM mass, and the redshift, and then provide semi-analytic expressions.
We also find that the core mass follows the log-normal distribution for a given halo mass, both in our model and in simulation results, and show that not all but a sizable fraction of the scatter in the CHMR found in the numerical simulations can be explained by the scatter of the concentration parameter.

The rest of the paper is organized as follows.
In Sec.~\ref{sec:review}, we summarize the basic properties of the FDM model related to our study.
We use them to model the soliton core for a given halo mass and discuss the dependence of the CHMR on parameters and the distribution function of the core mass in Sec.~\ref{sec:model}.
Discussion and summary are given in Sec.~\ref{sec:summary_disscusion}.
Throughout the present paper, the physical size of a halo is defined as its virial radius, and we use the concentration-halo mass relation for the CDM halos given by \citet{2021MNRAS.506.4210I}. 
We also use the dimensionless Hubble parameter with $h=0.7$, the present matter density $\Omega_{\rm m0}=0.30$, and dark energy density $\Omega_{\rm \Lambda 0}=0.70$.

\section{Review of structure formation in the FDM model} \label{sec:review}
In this section, we review the basic equations of FDM and the previous studies related to the FDM halo properties.
We first review the basic equations in Sec.~\ref{subsec:review_basic_eq}, and the non-linear properties in Sec.~\ref{subsec:review_fdm_halo}.
One of the key properties to determine the structure of the FDM halos is the CHMR, and extensive studies are conducted in the literature as summarized in Sec.~\ref{subsec:review_previous_study}.

\subsection{Basic equations} \label{subsec:review_basic_eq}
A popular FDM candidate is a scalar particle minimally coupled to gravity whose mass is around $mc^{2} \simeq 10^{-23} - 10^{-21}\ {\rm eV}$.
The corresponding de Broglie wavelength $\lambda_{\rm dB}$ is
\begin{equation}
    \frac{\lambda_{\rm dB}}{2\pi} = \frac{\hbar}{mv} = 1.92\ {\rm kpc} \left(\frac{mc^{2}}{10^{-22}\ {\rm eV}}\right)^{-1} \left(\frac{v}{10\ {\rm km/s}}\right)^{-1},
\end{equation}
where $\hbar$ is the Dirac constant and $v$ is the velocity.
Interestingly, the above de Broglie wavelength is close to the typical length scale where the so-called small-scale problems arise.
The de Broglie wavelength is larger than the intraparticle distances of dark matter, and thus the wavefunctions spatially overlap, to
make the FDM exist in a coherent state.

The coherent state of FDM satisfies the Schr\"{o}dinger-Poisson (SP) equation \citep{1990PhRvD..42..384S,1993ApJ...416L..71W},
\begin{eqnarray}
    &&i\hbar\frac{\partial \psi}{\partial t} = -\frac{\hbar^{2}}{2m}\nabla^{2}\psi + m\Phi\psi \label{SP_1},\\
    &&\nabla^{2} \Phi = 4\pi G m |\psi|^{2} \label{SP_2},
\end{eqnarray}
where $\psi$ is the complex wavefunction of the coherent state, $G$ is the Newtonian constant of gravitation, and $\Phi$ is the gravitational potential.
Note that the variables including spatial coordinates $\boldsymbol{x}$ and time $t$ are physical quantities; we ignore the cosmological expansion in this study because we only consider gravitationally bound objects.
From the SP equation, we may understand that the system is described by quantum mechanics with gravitational potential induced by itself.
The SP equation is invariant under the transformation:
\begin{align}
    &\{x,t,\rho,m,\psi,\Phi, M\}\hspace{45mm}& \nonumber\\
    &\rightarrow \{\alpha x,\beta t, \beta^{-2}\rho, \alpha^{-2} \beta m, \alpha \beta^{-3/2}\psi, \alpha^{2}\beta^{-2}\Phi, \alpha^{3}\beta^{-2}M\},& \label{sp_scale_sym}
\end{align}
where $\rho = m |\psi|^2$ is the mass density profile and $M = \int d^3 x \rho$ is the enclosed mass \citep{1990PhRvD..42..384S}.
The invariance is an interesting feature of the soliton core as discussed in Sec.~\ref{subsec:review_fdm_halo}.

The SP equation can be derived from the Heisenberg equation of motion with the second-quantized $N$-body Hamiltonian by applying Bogoliubov's prescription or the mean-field approximation,
\begin{equation}
    \hat{\Psi}(\boldsymbol{x}, t) = \langle\hat{\Psi}(\boldsymbol{x}, t)\rangle + \delta \hat{\Psi}(\boldsymbol{x}, t) \equiv \psi(\boldsymbol{x}, t) + \delta \hat{\Psi}(\boldsymbol{x}, t),
\end{equation}
where $\hat{\Psi}$ is the Bose operator describing the many-body system.
The wavefunction $\psi$ describes a classical field with a non-zero ensemble average taken in terms of the second-quantization.
Note that $\psi$ is the first-quantized wavefunction, whose wave nature is incorporated in the SP equation \citep{2021A&ARv..29....7F}.

It is useful to express the SP system as a fluid in order to gain physical insight.
Let us first decompose the wavefunction as 
\begin{equation}
    \psi = |\psi| e^{i\theta} = \sqrt{\frac{\rho}{m}}e^{i\theta} \label{decompose_psi_rho}.
\end{equation}
We may relate the gradient of the phase to velocity as
\begin{equation}
    \boldsymbol{v} = \frac{\hbar}{m}\nabla{\theta} \label{decompose_psi_velocity}.
\end{equation}
Then, substituting Eqs.~\ref{decompose_psi_rho} and \ref{decompose_psi_velocity} into Eqs.~\ref{SP_1} and \ref{SP_2}, we can obtain the fluid representation, {\it{i.e.,}} the Madelung equation \citep{1927ZPhy...40..322M},
\begin{eqnarray}
    &&\frac{\partial \rho}{\partial t} + \nabla \cdot (\rho \boldsymbol{v}) = 0,\\
    &&\frac{\partial \boldsymbol{v}}{\partial t} + (\boldsymbol{v}\cdot \nabla)\boldsymbol{v} = -\nabla \Phi + \frac{\hbar^{2}}{2m^{2}} \nabla \left(\frac{\nabla^{2}\sqrt{\rho}}{\sqrt{\rho}}\right)\label{fdm_Euler}.
\end{eqnarray}
The last term in Eq.~\ref{fdm_Euler} is called the quantum pressure term,
which originates from the uncertainty principle. The quantum pressure
effectively suppresses structure formation at small length scales.

Although we can easily see how the quantum nature comes into play in FDM halos in the Madelung equation, it is not fully consistent with the SP equation since the velocity field in 
Eqs.~\ref{decompose_psi_rho} and \ref{decompose_psi_velocity} is ill-defined
in regions where the density equals to zero.
Equivalently, the Madelung equation only considers a fluid with non-rotating velocity.
As shown in \citet{1994PhRvA..49.1613W}, additional quantization conditions are needed to obtain the SP equation from the Madelung equation.
FDM simulations based on the Madelung equation ({\it{e.g.,}} \citet{2018PhRvD..97h3519M}) might 
not properly treat small-scale features.

It is important to note that, while Eq.~\ref{fdm_Euler} is similar to the Jeans equation obtained from the Boltzmann equation, the nature of the pressure term is different. 
The pressure term in the Jeans equation obtained from the Boltzmann equation is determined by the local velocity dispersion as well as the density,
but the quantum pressure term can be calculated for a given density distribution.
In other words, the former needs an additional equation, {\it{i.e.,}} an effective equation of state, to determine the evolution of the system, while the latter gives a closed form.
The quantum pressure comes from the divergence of the stress tensor $\boldsymbol{\sigma}$,
\begin{eqnarray}
    &&\boldsymbol{\nabla} \cdot \boldsymbol{\sigma} = \rho \cdot \frac{\hbar^{2}}{2m^{2}} \nabla \left(\frac{\nabla^{2}\sqrt{\rho}}{\sqrt{\rho}}\right)\\
    &&\sigma_{ij} = \frac{\hbar^{2}\rho}{4m^{2}} \frac{\partial^{2} \log \rho}{\partial x_{i}\partial x_{j}}.
\end{eqnarray}
In the linear theory of structure formation, the Jeans scale, above which density perturbations are gravitationally unstable, can be obtained from Eq.~\ref{fdm_Euler} as \citet{2000PhRvL..85.1158H}, 
\begin{align}
    k_{\rm J} &=(16 \pi G m^2 \rho_{\rm m0} \,a)^{\frac{1}{4}} \nonumber \\ &=70(1+z)^{-\frac{1}{4}}
    \left(\frac{\Omega_{\rm m0}}{0.3}\right)^{\frac{1}{4}}
    \left(\frac{H_{0}}{70\ {\rm km\ s^{-1}\ Mpc^{-1}}}\right)^{\frac{1}{2}} \nonumber
    \\ & \quad \times 
    \left(\frac{mc^{2}}{10^{-22}\ {\rm eV}}\right)^{\frac{1}{2}}\ {\rm Mpc^{-1}},
\end{align}
where $\rho_{\rm m0}$, $a$, $z$, and $H_0$ are the present mean dark matter density, the expansion parameter, cosmological redshift, and the Hubble constant, respectively.
The Jeans mass $M_{\rm J}$ can be estimated as
\begin{align}
    M_{\rm J} &= 10^{7}(1+z)^{\frac{3}{4}}
    \left(\frac{\Omega_{\rm m0}}{0.3}\right)^{\frac{1}{4}}
    \left(\frac{H_{0}}{70\ {\rm km\ s^{-1}\ Mpc^{-1}}}\right)^{\frac{1}{2}} \nonumber
    \\ & \quad \times
    \left(\frac{mc^{2}}{10^{-22}\ {\rm eV}}\right)^{-\frac{3}{2}}\ M_{\odot} \label{fdm_jeans_mass}.
\end{align}
Structure formation is effectively suppressed below the Jeans mass.

\subsection{Properties of the FDM halo} \label{subsec:review_fdm_halo}
The structure of FDM halos has been studied using cosmological simulations, which solve the SP equation (Eqs.~\ref{SP_1} and \ref{SP_2}) ({\it{e.g.,}} \citet{2014NatPh..10..496S, 2021MNRAS.506.2603M}).
It is known that the soliton core exists in the center of each halo surrounded by an outskirt component.
More precisely, a FDM halo consists of many granular structures that are generated by the wave interference, and the spatially averaged structure resembles the NFW density profile (see modeling of granular structures in \citet{2022ApJ...925...61K}).
The FDM density profile can be expressed as
\begin{equation} \label{FDMprofile}
    \rho(r) = 
    \begin{cases}
        \rho_{\rm sol}(r) & \text{$r < r_{\rm t}$}  \\
        \rho_{\rm NFW}(r) & \text{$r > r_{\rm t}$}, 
    \end{cases}
\end{equation}
where $r_{\rm t}$ is the transition radius between the NFW profile and the soliton core 
with $\rho_{\rm sol}(r_{\rm t})  = \rho_{\rm NFW}(r_{\rm t})$.
The soliton core profile can be expressed by the ground state solution of the SP equation.
An empirical form is given by \citet{2014NatPh..10..496S} as
\begin{equation}
    \rho_{\rm sol}(r) = \frac{\rho_{\rm c}}{\{1+0.091(r/r_{\rm c})^{2}\}^{8}}, \label{soliton_emp}
\end{equation}
where 
\begin{equation}
    \rho_{\rm c} = 0.019\left(\frac{mc^{2}}{10^{-22}\ {\rm eV}}\right)^{-2} \left(\frac{r_{\rm c}}{\rm kpc}\right)^{-4}\ M_{\odot}\ \rm{pc}^{-3}. \label{rho_c_dep}
\end{equation}
Here $r_{\rm c}$ represents the core radius where the density drops to half of its central density.
Since the soliton core profile decays quickly at $r_\mathrm{c}$, the transition radius $r_{\rm t}$ is estimated to be roughly a few times the core radius $r_{\rm c}$, which is also indicated by numerical simulations.
The core mass $M_{\rm c}$ is defined as the enclosed mass within the core radius $r_{\rm c}$ and it can be expressed as
\footnote{In the following, 
\begin{equation}
    M_\mathrm{x}(<r) \equiv \int_0^r dr' 4 \pi r'^2 \rho_\mathrm{x}(r') \label{M_c_dep}
\end{equation}
denotes the enclosed mass for the profile ``x'' (= ``sol'' or ``NFW'') within a radius $r$. 
}
\begin{eqnarray}
    M_{\rm c} &=& M_{\rm sol} (< r_{\rm c}) \nonumber \\
    &=& 5.3 \times 10^{7} \left(\frac{mc^{2}}{10^{-22}\ {\rm eV}}\right)^{-2} \left(\frac{r_{\rm c}}{{\rm kpc}}\right)^{-1}\ M_{\odot} \label{rctrue_Mc}.
\end{eqnarray}
The parameter dependence in Eqs.~\ref{soliton_emp} and \ref{rctrue_Mc} is consistent with the scale invariance of the SP equation in Eq.~\ref{sp_scale_sym}.
Note that the core radius is determined once we determine the core mass.
The soliton core profile is then determined with a single parameter for a given FDM mass $m$, and we choose the core mass $M_{\rm c}$ as a quantity that 
characterizes the soliton core in this study.

The NFW profile can be expressed as
\begin{equation}
    \rho_{\rm NFW}(r) = \frac{\rho_{\rm s}}{(r/r_{\rm s})(1+r/r_{\rm s})^{2}}, \label{nfw}
\end{equation}
where $r_{\rm s}$ and $\rho_{\rm s}$ are the scale radius and density, respectively.
Instead of $r_{\rm s}$ and $\rho_{\rm s}$, the total halo mass $M_{\rm h}$ and the concentration parameter $c_{\rm vir}$, which are functions of $r_{\rm s}$ and $\rho_{\rm s}$, are widely used to describe the nature of the NFW profile. 
The concentration parameter is defined as $c_{\rm vir} \equiv r_{\rm vir}/r_{\rm s}$, where $r_{\rm vir}$ is the virial radius of the halo.
Note that the definition of the concentration parameter is different according to the different definitions of the halo radius.
In this paper, we use the virial radius as the definition for the halo radius as the literature we refer to in Sec.~\ref{subsec:review_previous_study}.
The enclosed mass of the NFW profile is given as 
\begin{equation}
    M_{\rm NFW}(<r) = 4\pi \rho_{\rm s} r_{\rm s}^{3}\left\{\ln\left(1+\frac{r}{r_{\rm s}}\right) - \frac{r/r_{\rm s}}{1 + r/r_{\rm s}}\right\}. \label{NFW_enclosed_1}
\end{equation}
Using this equation, the halo mass can be written as
\begin{equation}
    M_{\rm h} = M_{\rm NFW}(<r_{\rm vir}) = 4\pi \rho_{\rm s} r_{\rm s}^{3}f(c_{\rm vir}), \label{NFW_enclosed}
\end{equation}
where $f(c_{\rm vir})$ is defined as $f(c_{\rm vir}) \equiv \ln(1+c_{\rm vir}) - c_{\rm vir}/(1+c_{\rm vir})$.
The virial halo mass can be defined via the mean matter density $\rho_{\rm m}$ at a given redshift $z$ as
\begin{equation}
    M_{\rm h} = \frac{4}{3}\pi r_{\rm vir}^{3} \zeta(z) \rho_{\rm m}(z) \label{rvir_def},
\end{equation}
where $\zeta(z)$ is defined as
\begin{equation}
    \zeta(z) = \frac{1}{\Omega_{\rm m}(z)}\{18\pi^{2} + 82(\Omega_{\rm m}(z)-1) - 39(\Omega_{\rm m}(z)-1)^{2}\}, \label{def_zeta}
\end{equation}
with $\Omega_{\rm m}(z)$ being the matter density parameter at redshift $z$ \citep{1998ApJ...495...80B}.

It is known that there is a relation between the concentration and the halo mass, known as $c_{\rm vir}$-$M_{\rm h}$ relation.
The $c_{\rm vir}$-$M_{\rm h}$ relation is studied by a number of cosmological CDM simulations ({\it{e.g.,}} \citet{2001MNRAS.321..559B, 2021MNRAS.506.4210I}).
\citet{2001MNRAS.321..559B} show that the mean concentration of halos with  mass $M_{\rm h}$ at redshift $z$ can be fitted well by
\begin{equation}
    c_{\rm vir}^{\rm B}(M_{\rm h}, z; {\rm CDM}) = A \frac{1+z_{\rm coll}(M_{\rm h}; P_{\rm CDM})}{1+z},
\end{equation}
where $P_{\rm CDM}$ is the linear matter power spectrum at redshift $z=0$, and the constant $A$ is chosen as 3.13 \citep{2015MNRAS.454.1958M}. 
The $z_{\rm coll}$ is the collapsing time obtained by
\begin{equation}
    D(z_{\rm coll}) \sigma(f_{\rm coll}M_{\rm h}; P_{\rm CDM}) = \delta_{c},
\end{equation}
where $D(z)$ is the linear growth rate and $\sigma(M)$ is the linear root-mean-square density fluctuation on the comoving scale encompassing a mass $M$. 
The constant $f_{\rm coll}$ is set to $0.01$ and $\delta_{c}$ represents the critical density contrast for collapse, $\delta_{c} = 1.59 + 0.0314 \ln \sigma_{8}(z)$.
\citet{2021MNRAS.506.4210I} is the cosmological CDM simulation with the largest box size of $140\ h^{-1}{\rm Mpc}$.
The $c_{\rm vir}$-$M_{\rm h}$ relation obtained in \citet{2021MNRAS.506.4210I} is roughly consistent with the analytic estimate above.
The dependence of the concentration parameter on the halo mass at redshift $z=0$ in the CDM model is roughly
\begin{eqnarray}
    &&c_{\rm vir} \propto M_{\rm h}^{-0.06}\ \ \ \ \ \ \ \ \text{$M_{\rm h} \lesssim 10^{11}\ h^{-1} M_{\odot}$}\label{cMh_cdm_low} \\
    &&c_{\rm vir} \propto M_{\rm h}^{-0.12}\ \ \ \ \ \ \ \ \text{$M_{\rm h} \gtrsim 10^{11}\ h^{-1}M_{\odot}$} \label{cMh_cdm_high}.
\end{eqnarray}
By using the scale radius and the scale density following $r_{\rm s} \propto c_{\rm vir}^{-1} M_{\rm h}^{1/3}$ and $\rho_{\rm s} \propto c_{\rm vir}^{3}f^{-1}(c_{\rm vir})$ at a given redshift, respectively, which can be obtained by Eqs.~\ref{NFW_enclosed} and \ref{rvir_def}, we can also estimate the halo mass dependence of them.
We can obtain the CDM concentration parameter using the python package COLOSSUS \citep{2018ApJS..239...35D}.
Here we use the default cosmological parameters in COLOSSUS with version 1.2.17, which might be slightly different from our choice.
Note that the concentration has a substantial scatter around the mean value.
For example, \citet{2015ApJ...799..108D} find a $1 \sigma$ scatter of about $0.16$ dex nearly independent of the halo mass and redshift.

The $c_{\rm vir}$-$M_{\rm h}$ relation in the FDM halos has been studied, recently in \citet{2022MNRAS.515.5646D} and \citet{2022MNRAS.517.1867L}.
\citet{2022MNRAS.515.5646D} refers to the case of the warm dark matter universe studied in \citet{2012MNRAS.424..684S}, showing that the concentration parameter in the FDM halos are expressed as
\begin{eqnarray}
    &c_{\rm vir}(M_{\rm h}, z; {\rm FDM})\hspace{50mm}& \nonumber \\
    &= c_{\rm vir}^{\rm B}(M_{\rm h}, z; {\rm FDM}) \Delta^{\rm FDM}(M_{0}, \gamma_{0}, \gamma_{1}, \gamma_{2}),& 
\end{eqnarray}
where the additional suppression factor $\Delta^{\rm FDM}$ is defined by
\begin{align}
    &\Delta^{\rm FDM}(M_{0}, \gamma_{0}, \gamma_{1}, \gamma_{2}) \nonumber \\
    &\hspace{5mm} = \left(1+\frac{M_{0}}{f_{\rm coll}M_{\rm h}}  \right)^{-\gamma_{0}} \left(1+\gamma_{1} \frac{M_{0}}{M_{\rm h}} \right)^{-\gamma_{2}},
\end{align}
where $\gamma_{0} = d\ln c_{\rm vir}^{\rm B}/d\ln M_{\rm h}|_{M_{\rm h}= 4M_{0}}$, $\gamma_{1} = 15$, $\gamma_{2} = 0.3$, and 
\begin{equation}
    M_{0} = 1.6\times 10^{10}\ M_{\odot} \left(\frac{mc^{2}}{10^{-22}\ \rm{eV}}\right)^{-\frac{4}{3}}.
\end{equation}
The linear matter power spectrum of FDM models, $P_{\rm FDM}$, which is needed to calculate $c_{\rm vir}^{\rm B}(M_{\rm h}, z; {\rm FDM})$, is given as
\begin{equation}
    P_{\rm FDM}(k) = T^{2}_{\rm F}(k) P_{\rm CDM}(k), \ \ T_{\rm F}(k) \simeq \frac{\cos x^{3}}{1+x^{8}}, \label{transf}
\end{equation}
with $x = 1.61 (mc^{2}/10^{-22}\ {\rm eV})^{1/18}\ k/k_{\rm J}$ as shown by \citet{2000PhRvL..85.1158H}.
\citet{2015MNRAS.451.3117S} show a turnover feature of the $c_{\rm vir}$-$M_{\rm h}$ relation for models with suppressed small-scale power. 
Based on this, \citet{2022MNRAS.517.1867L} obtains the following simple expression;
\begin{equation}
    c_{\rm vir}(M_{\rm h}, z; {\rm FDM}) = c_{\rm vir}(M_{\rm h}, z; {\rm CDM})\ F\left(\frac{M_{\rm h}}{M_{\rm h}^{\rm hm}}\right), \label{cfdm_def}
\end{equation}
where function $F$ is defined as $F(x) = (1+ax^{b})^{c}$ with $(a,b,c) = (5.496, -1.648, -0.417)$
\footnote{It seems that there is a typo in the main text in \citet{2022MNRAS.517.1867L}. We use the Fig.~1 in their paper to obtain these values.}
, and $M_{\rm h}^{\rm hm}$ is the half-mode mass,
\begin{equation}
    M_{\rm h}^{\rm hm} = 3.8\times 10^{10}\ M_{\odot} \left(\frac{mc^{2}}{10^{-22}\ \rm{eV}}\right)^{-\frac{4}{3}} \label{half_mode_mass}.
\end{equation}
The $c_{\rm vir}$-$M_{\rm h}$ relation obtained in \citet{2022MNRAS.517.1867L} has a steeper turnover than that of \citet{2022MNRAS.515.5646D}, and the concentration parameter becomes $c_{\rm vir} < 1$ in halos with the Jeans mass. 
The turnover occurs around $M_\mathrm{h} \simeq 4 M_{\rm h}^{\rm hm} (\equiv M_{\rm h}^{\rm 4hm})$, four times larger than the half-mode mass.
Note that the $c_{\rm vir}$-$M_{\rm h}$ relation obtained in \citet{2022MNRAS.515.5646D} can be expressed in the same way as \citet{2022MNRAS.517.1867L}. 
We find $F(x)$ with $(a,b,c) = (9.431, -1.175, -0.232)$ can reproduce the FDM suppression of the $c_{\rm vir}$-$M_{\rm h}$ relation obtained by \citet{2022MNRAS.515.5646D}.
In the following, when we refer to \citet{2022MNRAS.515.5646D}, we use this expression rather than the original one.

\begin{figure}
    \includegraphics[width=\columnwidth]{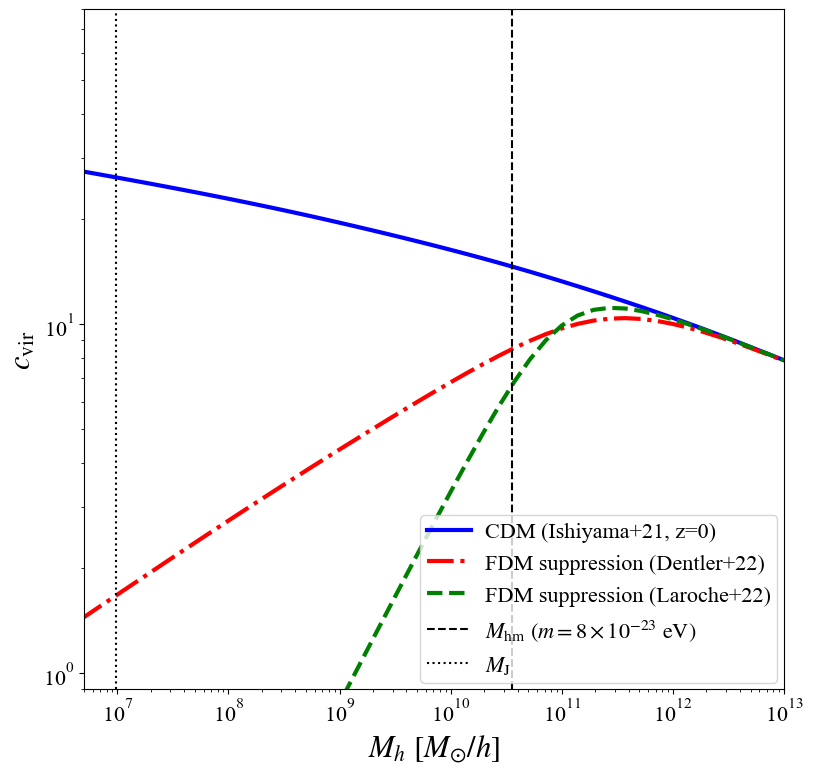}
    \caption{The suppression of the $c_{\rm vir}$-$M_{\rm h}$ relation in the FDM halos at redshift $z=0$.
    The blue solid line shows the $c_{\rm vir}$-$M_{\rm h}$ relation obtained in the largest cosmological CDM simulation \citet{2021MNRAS.506.4210I}, which approximately follows $c_{\rm vir} \propto M_{\rm h}^{-0.06}$ for $M_{\rm h} \lesssim 10^{11}\ h^{-1}M_{\odot}$ and $c_{\rm vir} \propto M_{\rm h}^{-0.12}$ for $M_{\rm h} \gtrsim 10^{11}\ h^{-1}M_{\odot}$.
    The red dash-dotted and green dashed lines consider the FDM suppression factors obtained by \citet{2022MNRAS.515.5646D} and \citet{2022MNRAS.517.1867L}, respectively.
    The shape of the $c_{\rm vir}$-$M_{\rm h}$ relation in the FDM halos has turnover around $M_{\rm h} \simeq M_{\rm h}^{\rm 4hm}$.
    For the FDM halos with $M_{\rm h} < M_{\rm h}^{\rm 4hm}$, the relation is $c_{\rm vir} \propto M_{\rm h}^{0.21}$ in the case of \citet{2022MNRAS.515.5646D}. 
    \citet{2022MNRAS.517.1867L} shows sharper turnover and $c_{\rm vir} < 1$ in halos with the Jeans mass.
    }
    \label{fig:c-M}
\end{figure}

Fig.~\ref{fig:c-M} shows the $c_{\rm vir}$-$M_{\rm h}$ relation that we use in this study and comparison of the FDM suppression between \citet{2022MNRAS.515.5646D} and \citet{2022MNRAS.517.1867L}.
Note again that we use the $c_{\rm vir}$-$M_{\rm h}$ relation in the CDM halos studied in \citet{2021MNRAS.506.4210I} with the help of COLOSSUS.
In this study, we refer to the FDM suppression shown in \citet{2022MNRAS.515.5646D} to obtain the concentration parameter in the FDM halos since it seems that the concentration calculated by \citet{2022MNRAS.517.1867L} might be too small around the halo with the Jeans mass as we stated above.
Using Eqs.~\ref{cMh_cdm_low} and \ref{cMh_cdm_high}, and the suppression factor $F$, we can obtain the halo mass dependence on the concentration in the FDM halos as
\begin{eqnarray}
    &&c_{\rm vir} \propto M_{\rm h}^{0.21}\ \ \ \ \ \ \ \ \ \ \text{$M_{\rm h} \lesssim M_{\rm h}^{\rm 4hm}$}\label{cMh_fdm_low} \\
    &&c_{\rm vir} \propto M_{\rm h}^{-0.12}\ \ \ \ \ \ \ \ \text{$M_{\rm h} \gtrsim M_{\rm h}^{\rm 4hm}$}. \label{cMh_fdm_high}
\end{eqnarray}
Note that we also obtain the following fitting formulae, 
\begin{eqnarray}
    &&f(c_{\rm vir}) \propto M_{\rm h}^{0.22}\ \ \ \ \ \ \ \ \ \ \text{$M_{\rm h} \lesssim M_{\rm h}^{\rm 4hm}$}\label{fcMh_fdm_low} \\
    &&f(c_{\rm vir}) \propto M_{\rm h}^{-0.07}\ \ \ \ \ \ \ \ \text{$M_{\rm h} \gtrsim M_{\rm h}^{\rm 4hm}$}, \label{fcMh_fdm_high}
\end{eqnarray}
which we use in Sec.~\ref{subsec:Relax_analysis}.
We follow \citet{2015ApJ...799..108D} and assume the log-normal distribution of the concentration with the scatter of $0.16$ dex regardless of the halo mass and the redshift.

\subsection{Previous studies on the core-halo mass relation} \label{subsec:review_previous_study}
The FDM halo profile Eq.~\ref{FDMprofile} does not determine the relation between the core mass $M_{\rm c}$ and the total halo mass $M_{\rm h}$ as it is unless we determine the relation between the parameters in the soliton profile and the NFW profile. 
However, there should be some underlying physics that relates to the inner core profile and the outer NFW profile.
Indeed, the relation between $M_{\rm c}$ and $M_{\rm h}$, the CHMR, has been identified as one of the key quantities to understanding the structure of the FDM halos.
\citet{2014PhRvL.113z1302S} obtains the CHMR by fitting the simulation results and with an analytic estimate as
\begin{equation}
    (1+z)^{-\frac{1}{2}} M_{\rm c} = \frac{1}{4}\left(\sqrt{\frac{\zeta(z)}{\zeta(0)}}\frac{M_{\rm h}}{M_{\rm min,0}}\right)^{\frac{1}{3}} M_{\rm min,0},
\end{equation}
where 
\begin{align}
     M_{\rm min,0} &= 4.4\times 10^{7} \ M_{\odot} \left(\frac{mc^{2}}{10^{-22}\ {\rm eV}}\right)^{-\frac{3}{2}} \left(\frac{\zeta(0)}{337.1}\right)^{\frac{1}{4}} \nonumber \\
     & \quad \times \left(\frac{H_{0}}{70\ {\rm km\ Mpc^{-1}\ s^{-1}}}\right)^{-\frac{3}{2}} \left(\frac{\Omega_{\rm m0}}{0.3}\right)^{-\frac{3}{4}} \nonumber \\ 
     & \quad \times \left(\frac{\rho_{\rm m0}}{40.8\ M_{\odot}\ {\rm kpc^{-3}}}\right). \label{Schive_min_mass} 
\end{align}
This relation is a single power law ($M_{\rm c} \propto M_{\rm h}^{\alpha}$) with the exponent $\alpha = 1/3$.
Here they obtain this relation by considering that the core radius can be determined by the de Broglie wavelength of FDM, where they use the halo velocity dispersion which is proportional to $M_{\rm h}^{1/3}$ at the leading order.
However, the different FDM simulations predict different powers of the CHMR, such as $\alpha = 5/9$ \citep{2018PhRvD..97h3519M} and $\alpha = 9/10$ \citep{2022MNRAS.511..943C}.
Moreover, the largest FDM cosmological simulation \citet{2021MNRAS.506.2603M} also shows a large scatter in the CHMR.
These differences might be due to the different setups such as box size, but \citet{2022MNRAS.511..943C} claims that there might be a large variance concerning the CHMR in the FDM halos.

There are several studies to construct the density profile in the FDM halos and/or the CHMR (semi-) analytically ({\it{i.e.,}} without simulation).
\citet{2019MNRAS.483..289R} uses the empirical relation obtained by \citet{2014PhRvL.113z1302S} to calculate the core radius for a given halo mass and determines the density profile by applying the density continuity condition with a radius several times the core radius.
Here the ratio between the core radius and the transition radius is allowed to have the halo mass dependence. 
\citet{2022PhRvD.105j3506Z} and  \citet{2022PhRvD.105b3512Y} consider the time-averaged potential in the SP equation and decompose the FDM halo profile into eigenstates, showing their time evolution in FDM halos.
\citet{2022PhRvD.105b3512Y} considers the impact of the power $\alpha$ of the CHMR on the density profile.
\citet{2022PhRvD.106j3532T} considers the NFW potential as a background to solve the linearized SP equation and obtain the CHMR, showing that it cannot be expressed in a single power law.
The impact of the scatter in the $c_{\rm vir}$-$M_{\rm h}$ relation on the CHMR is also discussed. 
The dedicated studies as such exist in the literature, but, there are no (semi-)analytic models which can fully explain the scatter on the CHMR that we observe in the simulations.
In that sense, we still do not understand the CHMR, and therefore more physically motivated (semi-)analytic modeling of the FDM halo, which can explain the scatter, is needed.

\section{Core-halo mass relation} \label{sec:model}
In this section, we describe our physical model for the CHMR in the FDM halos.
For simplicity, we consider the spherical static system.
We describe how we model the CHMR in Sec.~\ref{subsec:our_model}. 
In Sec.~\ref{subsec:jeans_relax}, we describe two physically motivated considerations to determine the characteristic radius.
The results are shown in Sec.~\ref{subsec:main_result} and the detailed behavior of our model is studied in Sec.~\ref{subsec:Relax_analysis}.
We give a comparison to other studies in Sec.~\ref{subsubsec:compare_other_models}, where we show that the core mass follows the log-normal distribution for a given halo mass, both in our model and in simulation results, and compare the scatters of the CHMR quantitatively.

\subsection{Modeling the core-halo mass relation} \label{subsec:our_model}
The assumption in our model is that the soliton core forms through mass redistribution of the NFW profile due to the wave nature of FDM.
This assumption might not be accurate in describing the FDM systems, however, it might be partially consistent with the setup of the FDM cosmological simulations which use the CDM power spectrum as the initial condition \citep{2014NatPh..10..496S, 2021MNRAS.506.2603M}.
For a given halo mass $M_{\rm h}$ and a redshift $z$, we first use the $c_{\rm vir}$-$M_{\rm h}$ relation obtained by the largest cosmological CDM simulation \citep{2021MNRAS.506.4210I} and the FDM suppression factor obtained by \citet{2022MNRAS.515.5646D} to construct the NFW profile as we describe in the Sec.~\ref{subsec:review_fdm_halo}.
Although both of the FDM simulations are performed using the CDM initial conditions, they should capture the impact of linear power suppression during the evolution as they start from a sufficiently high redshift ($z=127$).
Since the linear power suppression is incorporated in the concentration, we use the suppressed $c_{\rm vir}$-$M_{\rm h}$ relation.
The distribution of the concentration around the mean, which is determined by the $c_{\rm vir}$-$M_{\rm h}$ relation, is assumed to be the log-normal with the 1$\sigma$ scatter of 0.16 dex for all halos regardless of the redshift \citep{2015ApJ...799..108D}.
Note again that we use python package COLOSSUS \citep{2018ApJS..239...35D} to calculate the mean $c_{\rm vir}$-$M_{\rm h}$.
We consider two different physically motivated conditions to obtain the characteristic radius $\Tilde{r}_{\rm c}$ as we describe in Sec.~\ref{subsec:jeans_relax}.
We allow $\mathcal{O}(1)$ discrepancies between the characteristic radius and the mass-matching radius, which we normalize by using a free parameter $p_{1}$; $r_{\rm m} = p_{1} \Tilde{r}_{\rm c}$.
In our model, we impose that $p_{1}$ is constant for all halos at a given redshift.
Here the parameter $p_{1}$ is introduced to account for the uncertainty in the definition to calculate the characteristic radius.
Since the soliton core profile can be expressed with a single parameter for a given FDM mass $m$, we can finally obtain the core mass $M_{\rm c}$ by solving the mass continuity equation around $r_{\rm m}$, $M_{\rm NFW}(<r_{\rm m})=M_{\rm sol}(<r_{\rm m})$, where the left and right-hand sides indicate the enclosed mass of the NFW and the soliton core profile, respectively.
These steps of calculation give us the CHMR in the FDM halos.

We chose $p_1$ to reproduce \citet{2014NatPh..10..496S} at low redshift $z<1$ and with \citet{2021MNRAS.506.2603M} at redshift $z=3$, since these simulations solve the SP equation in a cosmological volume. 
After determining $p_{1}$ by fitting, the core mass $M_{\rm c}$ can be determined with 4 parameters in our model: the halo mass $M_{\rm h}$, the redshift $z$, the degree of the deviation $n\sigma$ of the concentration parameter from the mean $c_{\rm vir}$-$M_{\rm h}$ relation ($c_\mathrm{vir}^{(n)} = c_\mathrm{vir}^{\rm mean}\times 10^{0.16 n}$), and the FDM mass $m$.

\subsection{Jeans model and Relaxation model} \label{subsec:jeans_relax}
We show the two physical considerations to calculate the characteristic radius $\Tilde{r}_{\rm c}$ in this subsection.
The first one is based on the hydrostatic equilibrium condition, and we call it the "Jeans model".
The other is based on a relaxation time condition, and we call it the "Relaxation model".

The idea of the Jeans model is that the characteristic radius is determined by the scale at which the pressure of a single granular structure is balanced by gravity in a halo satisfying the hydrostatic equilibrium condition \citep{2021PhRvD.103b3508L}.
It can be obtained by setting $\boldsymbol{v} = 0$ in the Jeans equation, Eq.~\ref{fdm_Euler}, giving 
\begin{equation}
    \frac{GM_{\rm sol} (< \Tilde{r}_{\rm c})}{\Tilde{r}_{\rm c}} \simeq \frac{\hbar^{2}}{m^{2}\Tilde{r}_{\rm c}^{2}} \label{Jeans_def}.
\end{equation}
By approximating $M_{\rm sol}(<\Tilde{r}_{\rm c}) = M_{\rm NFW}(<\Tilde{r}_{\rm c})$, we can determine the characteristic radius from the NFW profile by solving Eq.~\ref{Jeans_def}.
The characteristic radius in this model is the same as the de Broglie wavelength if we assume the Keplerian velocity $v = \sqrt{GM_{\rm NFW}(<r)/r}$ inside halos,
\begin{equation}
    \Tilde{r}_{\rm c} \simeq \frac{\lambda_{\rm dB}(\Tilde{r}_{\rm c})}{2\pi} \label{Jeans_def_deBroglie}.
\end{equation}

While the Jeans model is based on the idea of determining the length scale where the wave nature of FDM is important, it does not include the notion of the dynamical evolution of the FDM halos. 
We here give another model that takes into account the dynamical process, that is, the Relaxation model.
The idea of the Relaxation model is that the characteristic radius is determined where the relaxation time equals the halo age \citep{2017PhRvD..95d3541H}.
The relaxation time in the FDM system can be approximated as follows.
The FDM system can be expressed with a superposition of energy eigenstates, which consists of not only the ground state but also the excited states, resulting in the interference pattern (granular structures) that can be observed in the FDM halos.
The size of the granular structures can be estimated by the de Broglie wavelength, and these structures can be regarded as quasi-particles with effective mass $m_{\rm eff} \simeq 4\pi \rho (\lambda_{\rm dB}/2)^{3}/3$ where $\rho$ is local density.
We can therefore estimate the relaxation time by the two-body relaxation of the granular structures with the effective mass \citep{2008gady.book.....B}
\begin{equation}
    t_{\rm relax}(r) \simeq \frac{0.1N}{\ln N}\ t_{\rm cr}(r) \label{two_body_relax_time},
\end{equation}
where the number of particles $N$ is evaluated as $N \simeq M(<r)/m_{\rm eff}$, with $M(<r) \simeq 4 \pi \rho r^{3}/3$ being the enclosed halo mass within $r$. 
The crossing time can be written as $t_{\rm cr}(r) = r/v$. 
Substituting these into Eq.~\ref{two_body_relax_time}, we can estimate the relaxation time at radius $r$ as
\begin{align}
    t_{\rm relax}(r) & \simeq \frac{0.1}{10} \frac{m^{3}v^{2}r^{4}}{\pi^{3}\hbar^{3}} \nonumber\\
    & \simeq 0.3\ \mathrm{Gyr} \left(\frac{v}{100\ \rm{km\ s^{-1}}}\right)^{2} 
    \left(\frac{r}{5\ \rm{kpc}}\right)^{4} \nonumber\\
    & \quad \times \left(\frac{mc^{2}}{10^{-22}\ {\rm eV}}\right)^{3}. \label{two_body_relax_time_fdm}
\end{align}
Here we ignore the parameter dependence of $\ln N$ and approximate it as $\ln N \sim 10$ for simplicity. 
It is consistent up to a factor of 2-3 for the mass range considered in this paper, $M_{\rm h}= 10^{7 \div 13}\ h^{-1}M_{\odot}$, $mc^{2}=8 \times 10^{-24\div-22}\ {\rm eV}$.
Note that the $\ln N$ factor is originally ignored in the \citet{2017PhRvD..95d3541H}.
Considering that the soliton core, which is the ground state, forms around the center of halos as a result of relaxation, we determine the characteristic radius as the radius where the relaxation time equals the halo age,
\begin{equation}
    t_{\rm relax}(\Tilde{r}_{\rm c}) = t_{\rm age}, \label{Relax_def}
\end{equation}
where the right-hand side corresponds to the halo age.
In this paper, we assume that the halo age at redshift $z$ is the same as that of the universe.
Note again that we assume the circular velocity $v$.

\subsection{Result} \label{subsec:main_result}
In both models, once we obtain ${\Tilde r}_\mathrm{c}$, we set the matching radius with a parameter $p_1$ to find the core mass.
We determine the (redshift dependent) parameter $p_{1}$ by fitting our calculation with the CHMR obtained in \citet{2014NatPh..10..496S} at redshift $z<1$ and \citet{2021MNRAS.506.2603M} at redshift $z=3$.
We set the fiducial FDM mass to $mc^{2} = 8.0\times 10^{-23}\ {\rm eV}$, while the original simulations \citet{2014NatPh..10..496S} and \citet{2021MNRAS.506.2603M} are conducted with $mc^{2} = 7.5\times 10^{-23}\ {\rm eV}$ and $7.0\times 10^{-23}\ {\rm eV}$, respectively.
We convert the original data for the CHMR into the relation with the fiducial FDM mass by using the scaling relation of the SP equation (Eq.~\ref{sp_scale_sym}).
Both core and halo masses are transformed as $M \propto m^{-3/2}$, which can be obtained by setting $\beta = 1$ in Eq.~\ref{sp_scale_sym}.
Note that the time scale in the simulation does not change in this transformation.

\begin{figure}
    \includegraphics[width=\columnwidth]{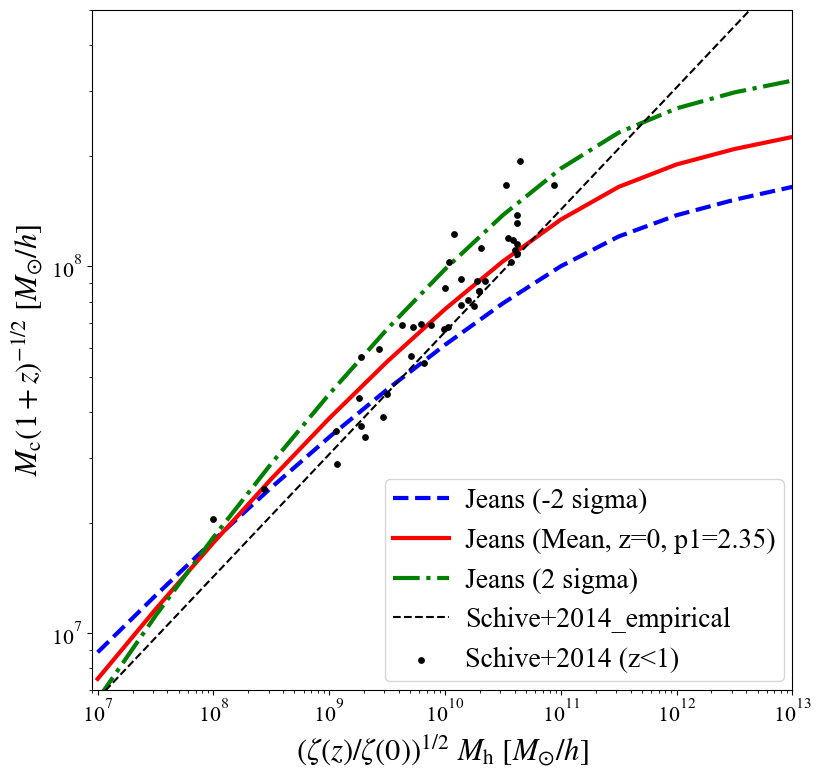}
    \caption{
    The comparison of the CHMR between the Jeans model at redshift $z = 0$ and the simulation data obtained by \citet{2014NatPh..10..496S} at redshift $z < 1$.
    We set the fiducial FDM mass $mc^{2} = 8.0 \times 10^{-23}\ {\rm eV}$.
    The red solid line shows the mean CHMR in the Jeans model at redshift $z = 0$ by using the mean $c_{\rm vir}$-$M_{\rm h}$ relation.
    The green dash-dotted and blue dashed lines show the relation when we consider the $2\sigma$ scatter in the $c_{\rm vir}$-$M_{\rm h}$ relation; $n=2, -2$, respectively.
    By setting $p_{1} = 2.35$, we successfully fit the CHMR  obtained by \citet{2014NatPh..10..496S} at redshift $z < 1$, which shows in black dots.
    The CHMR obtained from \citet{2014PhRvL.113z1302S} is plotted in a black dashed line.
    }
    \label{fig:jeans_z0}
\end{figure}

\begin{figure}
    \includegraphics[width=\columnwidth]{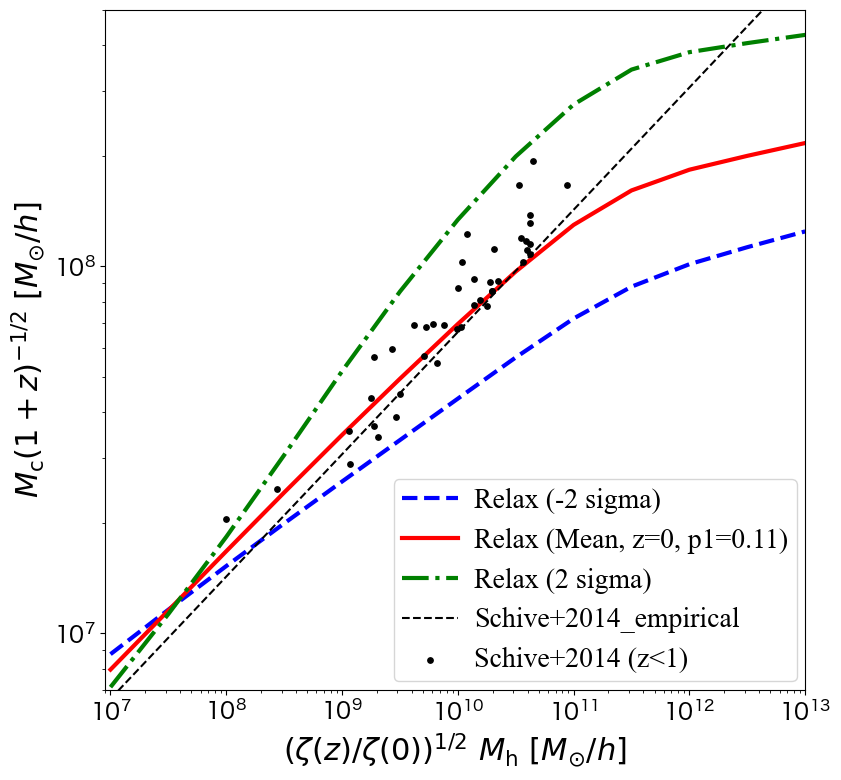}
    \caption{
    Similar to Fig.~\ref{fig:jeans_z0}, but for the Relaxation model.
    By setting $p_{1} = 0.11$, we successfully fit the CHMR obtained by \citet{2014NatPh..10..496S} at redshift $z < 1$.
    }
    \label{fig:relax_z0}
\end{figure}

Figure~\ref{fig:jeans_z0} shows the comparison of the CHMR between the Jeans model at redshift $z = 0$ and the simulation data obtained by \citet{2014NatPh..10..496S} at redshift $z < 1$.
Figure~\ref{fig:relax_z0} shows the same comparison, but we show the Relaxation model in this case.
The red solid lines in both figures show the mean CHMR in our models, in which we use the mean $c_{\rm vir}$-$M_{\rm h}$ relation.
We tune $p_{1}$ to fit the simulation data, which is shown in the black dots, finding $p_{1} = 2.35$ in the Jeans model and $p_{1} = 0.11$ in the Relaxation model are the best-fit values at low redshift $z < 1$.
Since the characteristic radius in the Jeans model corresponds to the de Broglie wavelength as shown in Eq.~\ref{Jeans_def_deBroglie}, it is considered reasonable for $p_{1}$ to be $2.0$-$3.0$, yielding the matching radius to be larger than de Broglie wavelength.
In the case of the Relaxation model, considering the definition where many granular structures relax and form the core at the center of the halo, the number of the granular structures within the characteristic radius is much larger than 1. 
Therefore, the characteristic radius should be much larger than the core (mass-matching) radius, and $p_{1}$ should be smaller than 1.
Both models show that the CHMR is roughly expressed in a double power law.
For the halo mass lower than $10^{11}\ h^{-1}M_{\odot}$, it is consistent with the empirical relation shown in \citet{2014PhRvL.113z1302S}, which are shown in the black dashed lines in the figures.
However, it is found that the power law index of the CHMR for more massive halos is smaller in both models.
The green dash-dotted and blue dashed lines consider the 2$\sigma$ scatter in the $c_{\rm vir}$-$M_{\rm h}$ relation, where 0.16 dex is taken as 1$\sigma$.
Here we show the 2$\sigma$ of the CHMR for reference to the degree of scatter.
We can see that three lines almost cross around the halo mass $M_{\rm h} \simeq 10^{7-8}\ h^{-1}M_{\odot}$, which we can interpret as the minimum halo mass as we show in Sec.~\ref{subsec:Relax_analysis}.
We also find that the more concentrated halos have more massive cores.
The scatter in the Relaxation model is larger than that in the Jeans model.
We can easily show this phenomenon by analyzing the Jeans model carefully as we do the Relaxation model in Sec.~\ref{subsec:Relax_analysis}, but we do not discuss it in this paper since it is not our main purpose.

\begin{figure}
    \includegraphics[width=\columnwidth]{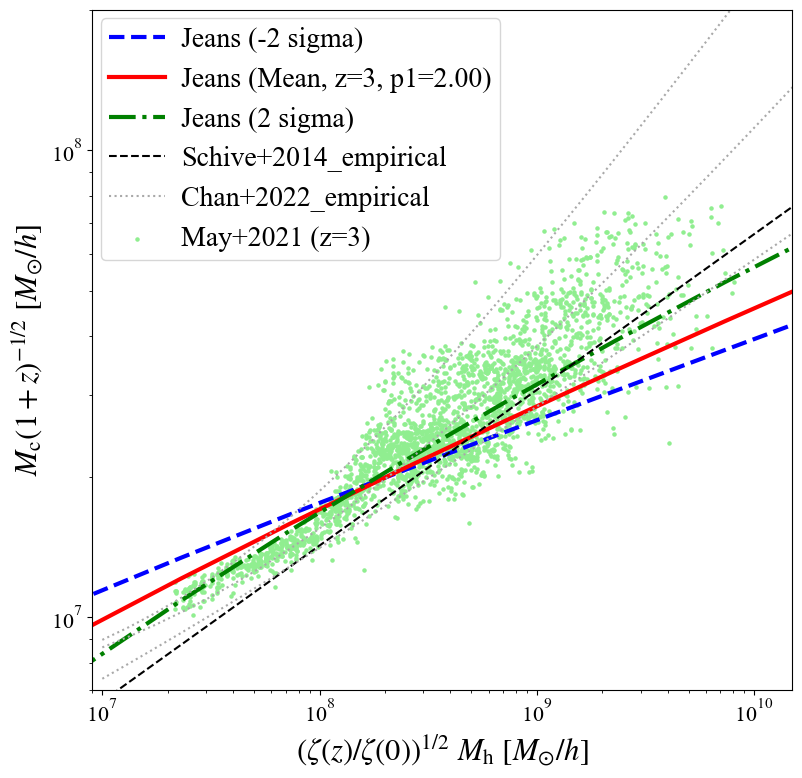}
    \caption{
    The comparison of the CHMR at redshift $z = 3$ between the Jeans model and the simulation data obtained by \citet{2021MNRAS.506.2603M}.
    We set the fiducial FDM mass $mc^{2} = 8.0 \times 10^{-23}\ {\rm eV}$.
    The red solid line shows the mean CHMR in the Jeans model at redshift $z = 3$ by using the mean $c_{\rm vir}$-$M_{\rm h}$ relation.
    The green dash-dotted and blue dashed lines show the relation when we consider the $2 \sigma$ scatter in the $c_{\rm vir}$-$M_{\rm h}$ relation; $n=2, -2$, respectively. 
    In this plot, we set $p_{1} = 2.00$.
    Although we try different $p_{1}$ values, the Jeans model cannot reproduce the estimated scatter of the CHMR obtained by \citet{2021MNRAS.506.2603M} at redshift $z = 3$, which shows in green dots.
    The estimated scatter in the CHMR empirically obtained by \citet{2022MNRAS.511..943C} is shown in grey dotted lines, and the black dashed line shows the empirical relation obtained by \citet{2014PhRvL.113z1302S}.
    }
    \label{fig:jeans_z3}
\end{figure}

\begin{figure}
	\includegraphics[width=\columnwidth]{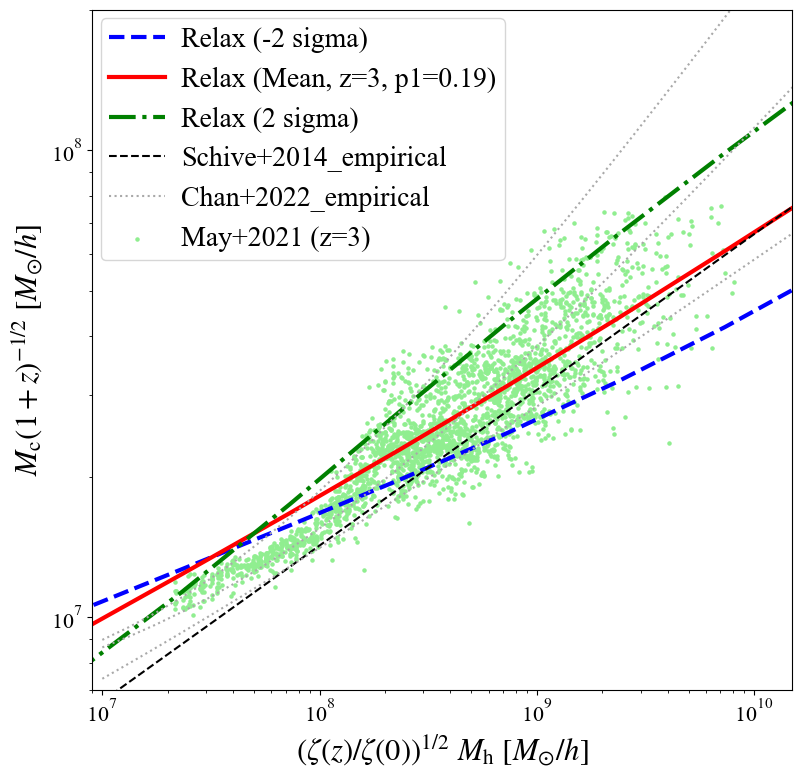}
    \caption{
    Similar to Fig.~\ref{fig:jeans_z3}, but for the Relaxation model. 
    In this case, by setting $p_{1} = 0.19$, it can reproduce the larger scatter and capture the similar trend including the minimum halo mass to the simulation data obtained by \citet{2021MNRAS.506.2603M} at redshift $z = 3$.
    }
    \label{fig:relax_z3}
\end{figure}

Figure~\ref{fig:jeans_z3} shows the comparison of the CHMR at redshift $z = 3$ between the Jeans model and the largest simulation data obtained by \citet{2021MNRAS.506.2603M}.
Figure~\ref{fig:relax_z3} shows the same comparison, but we show the Relaxation model in this case.
As in the same previous figures, the red solid, green dash-dotted, and blue dashed lines in both figures correspond to the mean, $n=2$, and $n=-2\sigma$ CHMR in our model, respectively, which we change the magnitude of the concentration. 
Again, we show the 2$\sigma$ of the CHMR for reference to the degree of scatter, and detailed analysis is shown in Sec.~\ref{subsubsec:compare_other_models}.
We allow the redshift dependence of $p_{1}$ and take $p_1=2.00$ for the Jeans model and 0.19 for the Relaxation model to reproduce the simulation data, which we plot in green dots.
Unfortunately, we do not find a better value of $p_{1}$ for the Jeans model that can reproduce the expected scatter in the CHMR.
Especially, the minimum halo mass, which is around the crossing point of three lines, seems to be different between the Jeans model and the simulation data.
On the other hand, in the case of the Relaxation model, it can reproduce the larger degree of the scatter than the Jeans model and it captures a similar trend to the simulation data including the minimum halo mass.
Note that the grey dotted lines are the estimated scatter of the CHMR shown by \citet{2022MNRAS.511..943C}, which is obtained by fitting the simulation data.

From these results, we conclude that the relaxation time condition Eqs.~\ref{two_body_relax_time_fdm} and \ref{Relax_def} is a more proper condition for the mass-matching radius to describe the FDM systems. 
Thus, in the following, we focus on the Relaxation model and understand its behavior.
The main goal is now to give the semi-analytic expression to the CHMR for the Relaxation model.

Before we end this subsection, we comment on the case of using the $c_{\rm vir}$-$M_{\rm h}$ relation for CDM halos.
As we mention in Sec.~\ref{subsec:our_model}, since the $c_{\rm vir}$-$M_{\rm h}$ relation calculated from the halos in the FDM simulations is thought to have the turnover around the half-mode mass, we use the $c_{\rm vir}$-$M_{\rm h}$ relation obtained by \citet{2022MNRAS.515.5646D}.
However, we also study the CHMR with the CDM $c_{\rm vir}$-$M_{\rm h}$ relation for comparison. 
In Figs.~\ref{fig:jeans_z0_cdm} to \ref{fig:relax_z3_cdm} in App.~\ref{CHMR_cdm}, we show the CHMR with the Jeans and the Relaxation models at redshift $z=0$ and $z=3$.
Although it seems that both models can reproduce the simulation data at redshift $z=3$ by setting $p_{1}=1.70$ for the Jeans model and $p_{1}=0.05$ for the Relaxation model, we find that the mean CHMR in both models with any $p_{1}$ value cannot explain the simulation data at $z<1$. 
From these results, we can understand that it seems better to use the $c_{\rm vir}$-$M_{\rm h}$ relation in FDM halos.
Once more, it is not surprising because the simulations are started at a sufficiently high redshift and should capture suppression of structure formation in the linear evolution.

\subsection{Analysis in the Relaxation model} \label{subsec:Relax_analysis}
As shown in Sec.~\ref{subsec:main_result}, the Relaxation model can reproduce the simulation data and show the larger scatter in the CHMR.
There are two important features in the Relaxation model. 
Namely, it is roughly expressed in a double power law, and the CHMRs with different magnitudes of the concentration intersect around the minimum mass.
In this section, we analyze the model at redshift $z=0$ to understand these features and express the CHMR in a simple form.
Though the same procedure can be applied to the analysis at different redshifts, we focus on redshift $z=0$ because we mainly need the information of low redshift to compare it with observations such as the rotation curves of galaxies and the gravitational lensing.
To obtain the CHMR in a simple form, we study the halo mass dependence of the characteristic radius and enclosed mass within the mass-matching radius.
We analyze the mean CHMR to show the double power law feature in Sec.~\ref{subsubsec:mean_analysis}.
The scatter in the CHMR and the minimum halo mass are discussed in Sec.~\ref{subsubsec:scatter_analysis}.
We discuss the FDM mass dependence in Sec.~\ref{subsubsec:fdm_mass_analysis} and show the semi-analytic expression of the CHMR at redshift $z=0$. 
Finally, we discuss the redshift dependence in Sec.~\ref{subsubsec:redshift_analysis}.

\subsubsection{Mean relation} \label{subsubsec:mean_analysis}
Let us remind us of the definition of the Relaxation model (Eqs.~\ref{two_body_relax_time_fdm} and \ref{Relax_def}), which can be rewritten as 
\begin{align}
    \Tilde{r}_{\rm c}^{3} M_{\rm NFW}(<\Tilde{r}_{\rm c}) &= 7 \times 10^{13} \left(\frac{mc^{2}}{10^{-22}\ {\rm eV}}\right)^{-3} \nonumber \\ & \quad \times \left(\frac{t_{\rm age}(z)}{13.8\ \rm Gyr}\right) \ M_{\odot}\ {\rm kpc}^{3} \label{relax_def_re}. 
\end{align}
Here we have taken $v$ to be the circular velocity.
Taking into account two cases according to whether the radius of concern is larger than the scale radius $r_{\rm s}$ or not, we approximate the enclosed mass of the NFW profile (Eq.~\ref{NFW_enclosed_1}) as
\begin{equation}
    M_{\rm NFW}(<r) \simeq 
    \begin{cases}
        2\pi \rho_{\rm s} r_{\rm s} r^{2} & \text{$r \ll r_{\rm s}$}  \\
        4\pi \rho_{\rm s}  r_{\rm s}^{3} \left\{\ln\left(\frac{r}{r_{\rm s}}\right) -1 \right\} & \text{$r \gg r_{\rm s}$}
    \end{cases}\label{nfw_enclosed_two_pattern}.  
\end{equation}
Then it can be expected that there are also two cases for the CHMR depending on whether the characteristic radius is larger than the scale radius, which would be the origin of the double power law feature.
Since $\Tilde{r}_{\rm c}$ monotonically decreases as the halo mass becomes larger (as we check later) and $r_{\rm s}$ behaves oppositely, the halo mass at the boundary of these two cases is uniquely determined, which can be estimated via the condition $\Tilde{r}_{\rm c} = r_{\rm s}$. 
Using the relation $r_\mathrm{vir} = c_{\mathrm{vir}} r_s$ and Eq.~\ref{rvir_def}, we obtain the boundary halo mass at a redshift $z$ as
\begin{align}
    M_{\rm h}^{\rm B} (z) &\simeq 1 \times 10^{11}\ M_{\odot} (1+z)^{\frac{3}{4}} \left(\frac{\zeta(z)}{\zeta(0)}\right)^{\frac{1}{2}} \left(\frac{mc^{2}}{10^{-22}\ {\rm eV}}\right)^{-\frac{3}{2}} \nonumber \\ 
    & \quad \times \left(\frac{c_{\mathrm{vir}}^{3}\{\ln(1+c_{\mathrm{vir}})-c_{\mathrm{vir}}/(1+c_{\mathrm{vir}})\}}{1000}\right)^{\frac{1}{2}} \nonumber \\
    & \quad \times \left(\frac{\rho_{\rm m0}}{40.8\ M_{\odot}\ {\rm kpc^{-3}}}\right)^{\frac{1}{2}}  \left(\frac{t_{0}}{13.8\ {\rm Gyr}}\right)^{\frac{1}{2}} \label{bound_mass}.
\end{align}
Here we approximate the halo age as $t_{\rm age}(z) \simeq t_{0}(1+z)^{-\frac{3}{2}}$ with $t_{0}$ being the age of the present universe, and we take $\Omega_{\rm m0} = 0.30$ to calculate $\zeta(0)$ defined in Eq.~\ref{def_zeta}.
Note that the right-hand side of Eq.~\ref{bound_mass} contains the concentration parameter, which is related to the halo mass.

We also have another typical halo mass that would affect the CHMR, that is, four times the half-mode mass at which the $c_{\rm vir}$-$M_\mathrm{h}$ relation shows turnover.
Thus, we would see the triple power law behavior in the CHMR, not the double power law discussed above.
However, since the boundary halo mass given in Eq.~\ref{bound_mass} at redshift $z=0$ is roughly the same as four times the half-mode mass for FDM mass $mc^{2} \simeq 10^{-23}-10^{-21}\ {\rm eV}$, 
\begin{equation}
    M_{\rm h}^{\rm B}(z=0) \simeq M_{\rm h}^{\rm 4hm},
\end{equation}
we do not consider the intermediate mass range between the boundary halo mass defined in Eq.~\ref{bound_mass} and $M_{\rm h}^{\rm 4hm}$ for $z=0$.
We only consider two cases in the following: halos with a mass larger/smaller than the boundary mass and four times the half-mode mass.
We call "low-mass halos" for halos with $M_{\rm h} < M_{\rm h}^{\rm 4hm}$, and "high-mass halos" for halos with $M_{\rm h} > M_{\rm h}^{\rm 4hm}$.

For the low-mass halos, $\Tilde{r}_{\rm c} \gtrsim r_{\rm s}$,  Eq.~\ref{relax_def_re} roughly becomes $\Tilde{r}_{\rm c}^{3} \ln (\Tilde{r}_{\rm c} / r_{\rm s}) \cdot 4\pi \rho_{\rm s} r_{\rm s}^{3} \simeq {\rm const.}$, where we ignore the subdominant term.
Remembering the halo mass dependence of the concentration parameter in this regime at redshift $z=0$, Eq.~\ref{fcMh_fdm_low}, and using Eq.~\ref{NFW_enclosed}, we obtain $4\pi \rho_{\rm s} r_{\rm s}^{3} \propto M_{\rm h}^{0.78}$.
Therefore, we obtain halo mass dependence of the characteristic radius, $\Tilde{r}_{\rm c} \propto M_{\rm h}^{-0.26 \div -0.20}$, where the uncertainty corresponds to the logarithmic factor.
To reproduce our numerical calculation, we find that $-0.21$ is the best-fit value for the power law index, which is consistent with this estimation.
For the high-mass halos, $\Tilde{r}_{\rm c} \lesssim r_{\rm s}$, Eq.~\ref{relax_def_re} becomes $\Tilde{r}_{\rm c}^{5} \cdot 2\pi \rho_{\rm s} r_{\rm s} = {\rm const.}$.
Using the halo mass dependence of the concentration parameter, Eqs.~\ref{cMh_fdm_high} and \ref{fcMh_fdm_high}, we obtain $2\pi \rho_{\rm s} r_{\rm s} \propto M_{\rm h}^{0.16}$ and $\Tilde{r}_{\rm c} \propto M_{\rm h}^{-0.032}$ through the same procedure.
To reproduce our numerical calculation, we find that $-0.06$ is the best-fit value for the power law index.
The possible reason for this difference is that the fitting region, $M_{\rm h} < 10^{13}\ h^{-1}M_{\odot}$, is not high enough to be valid to use the approximation of Eq.~\ref{nfw_enclosed_two_pattern}.
In summary, we obtain the following halo mass dependence of the characteristic radius:
\begin{eqnarray}
    &&\Tilde{r}_{\rm c} \propto M_{\rm h}^{-0.21}\ \ \ \ \ \ \ \ \text{low-mass halos} \label{tilde_rc_low_Mhdep}\\
    &&\Tilde{r}_{\rm c} \propto M_{\rm h}^{-0.06}\ \ \ \ \ \ \ \ \text{high-mass halos} \label{tilde_rc_high_Mhdep}.
\end{eqnarray}
We now see that $\Tilde{r}_{\rm c}$ monotonically decreases as a function of the halo mass, which validates our discussion above about the boundary halo mass, and the power law index changes around the boundary halo mass and four times the half-mode mass. 

Next, we study the halo mass dependence of the enclosed mass within the matching radius $r_{\rm m} = p_{1}\Tilde{r}_{\rm c}$ with $p_{1} = 0.11$.
From the definition of the Relaxation model Eq.~\ref{relax_def_re}, the relation $M_{\rm NFW}(<\Tilde{r}_{\rm c}) \propto \Tilde{r}_{\rm c}^{-3}$ holds.
If $p_{1} \simeq 1$, $M_{\rm NFW}(<\Tilde{r}_{\rm m}) \propto \Tilde{r}_{\rm m}^{-3}$ satisfies since $r_{\rm m} \simeq \Tilde{r}_{\rm c}$.
Even if the condition $p_{1} \simeq 1$ does not hold, this relationship is also valid if both of $\Tilde{r}_{\rm c}$ and $r_{\rm m}$ satisfy $\Tilde{r}_{\rm c}, r_{\rm m} \gg r_{\rm s}$ or $\ll r_{\rm s}$.
This is because the ratio between the enclosed halo mass within the characteristic radius and that within the matching radius can be estimated as $M_{\rm NFW}(<r_{\rm m})/M_{\rm NFW}(<\Tilde{r}_{\rm c}) = \ln (r_{\rm m}/r_{\rm s})/\ln (\Tilde{r}_{\rm c}/r_{\rm s})$ for the low-mass halos, and $(r_{\rm m}/\Tilde{r}_{\rm c})^{2}$ for the high-mass halos.
However, since $r_{\rm m} \simeq r_{\rm s}$ holds for the low-mass halos due to the small $p_{1}$ of 0.11, and $\Tilde{r}_{\rm c} \simeq r_{\rm s}$ holds for the high-mass halos as stated above, the relation deviates from this simple scaling.
We thus obtain the scaling by numerical fitting, finding $M_{\rm NFW}(<r_{\rm m}) \propto r_{\rm m}^{-2.2}$ for the low-mass halos and $M_{\rm NFW}(<r_{\rm m}) \propto r_{\rm m}^{-1.1}$ for the high-mass halos.
With the help of Eqs.~\ref{tilde_rc_low_Mhdep} and \ref{tilde_rc_high_Mhdep}, we determine the following halo mass dependence of the enclosed mass,
\begin{eqnarray}
    && M_{\rm NFW}(<r_{\rm m}) \propto M_{\rm h}^{0.46}\ \ \ \ \ \ \ \ \text{low-mass halos} \label{tilde_Mhrm_low_Mhdep}, \\
    && M_{\rm NFW}(<r_{\rm m}) \propto M_{\rm h}^{0.07}\ \ \ \ \ \ \ \ \text{high-mass halos}, \label{tilde_Mhrm_high_Mhdep}
\end{eqnarray}
which is also confirmed by numerical calculations.

\begin{figure}
\includegraphics[width=\columnwidth]{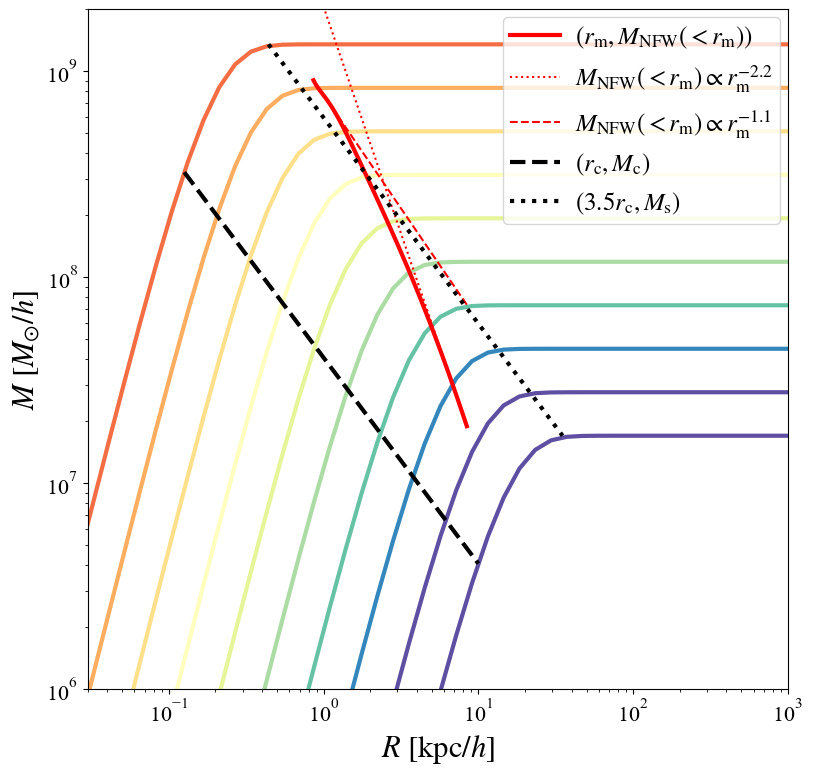}
    \caption{
    The enclosed masses with radius ($R=r$ or $r_\mathrm{m}$) are shown, which is used to determine the soliton core mass using the mass-matching condition.
    The red solid line shows the NFW enclosed mass at the matching radius $r_{\rm m}$ with halo mass range $M_{\rm h} = 10^{7-13}\ h^{-1}M_{\odot}$.
    The rainbow-colored lines show the enclosed mass of the soliton core with different core radii with the fixed FDM mass, $mc^{2} = 8\times 10^{-23}\ {\rm eV}$.
    For a given halo mass $M_\mathrm{h}$, we determine $(r_{\rm m}, M_{\rm MFW}(<r_{\rm m}))$ and then determine the core radius and the core mass such that the corresponding soliton core passes through the point.
    The red dotted line shows the power law in the low-mass halos, $M_{\rm MFW}(<r_{\rm m}) \propto r_{\rm m}^{-2.2}$, and the red dashed line shows the power law in the high-mass halos, $M_{\rm MFW}(<r_{\rm m}) \propto r_{\rm m}^{-1.1}$. 
    The relation between the core radius $r_{\rm c}$ and the core mass $M_{\rm c}$ is plotted in the black dashed line.
    The black dotted line shows the relation between the 3.5 $r_{\rm c}$ and the total core mass $M_{\rm s}$, which clearly shows the coincidence of the boundary mass (radius). 
    }
    \label{fig:Mmatch}
\end{figure}

Since we impose the mass continuity condition around the matching radius, $M_{\rm sol}(<r_{\rm m}) = M_{\rm NFW}(<r_{\rm m})$, to determine the soliton core profile, we next study the relation between the enclosed soliton mass within the matching radius $M_{\rm sol}(<r_{\rm m})$ and the core mass $M_{\rm c} = M_{\rm sol}(<r_{\rm c})$.
Figure~\ref{fig:Mmatch} shows the enclosed soliton core mass within a radius $R=r$, $M_{\rm sol}(<r)$ for different soliton core radii for the fixed FDM mass $mc^2 = 8 \times 10^{-23}$ eV (rainbow-colored lines), and $M_{\rm NFW}(<r_{\rm m})$ as a function of $R=r_{\rm m}$ at $M_\mathrm{h}=10^{7-13}\ h^{-1}M_{\odot}$ (red solid line).
As shown before, higher-mass halos correspond to the smaller matching radii.
Since $r_\mathrm{m}$ is uniquely determined with the halo mass $M_\mathrm{h}$, the soliton core profile is determined by the line that passes the point ($r_{\rm m}, M_{\rm NFW}(<r_{\rm m})$).
With this procedure, we determine the soliton core mass for a given halo mass. 
The behavior of the rainbow-colored lines can be understood by the fact that it can be approximated as
\begin{eqnarray}
    M_{\rm sol}(<r) &\simeq&
    \begin{cases}
        \frac{4}{3}\pi \rho_{\rm c} r^{3}\ \propto m^{6}M_{\rm c}^{4} & \text{($r < 0.5r_{\rm c}$)} \\
        M_{\rm s}\ \ \ \ \ \ \ \ \ \propto M_{\rm c} & \text{($r > 3.5r_{\rm c}$)}
    \end{cases},  \label{enclosed_soliton_twocase}
\end{eqnarray}
where the total core mass is denoted by $M_{\rm s}$. 
We have used Eqs.~\ref{rho_c_dep} and \ref{rctrue_Mc} to determine the parameter dependence.
Since the density profile of the soliton core drops rapidly, the enclosed mass of the soliton core within the radius larger than 3.5 times the core radius is almost $M_{\rm s}$.

To understand the halo mass dependence of the core mass, we divide it into two cases according to whether the matching radius $r_{\rm m}$ is larger than $3.5 r_{\rm c}$ or not. 
Since the boundary halo mass of these two cases (where the red solid line crosses the black dotted line in Fig.~\ref{fig:Mmatch}) is roughly the same as the boundary mass between the low-mass halos and the high-mass halos that we have been using (where the red solid line changes its power law index), we do not need additional division. 
For the high-mass halos, where $r_{\rm m} > 3.5 r_{\rm c}$, the relation $M_{\rm NFW}(<r_{\rm m}) = M_{\rm s} \propto M_{\rm c}$ is satisfied and we finally obtain the halo mass dependence, $M_{\rm c} \propto M_{\rm h}^{0.07}$.
For the low-mass halos, it is difficult to estimate exactly since the matching radius is between $0.5r_{\rm c}$ and $3.5r_{\rm c}$.
Although it is not accurate, we can roughly estimate the halo mass dependence of the core mass by using the upper relation in Eq.~\ref{enclosed_soliton_twocase}, Eq.~\ref{tilde_Mhrm_low_Mhdep},
and the relation $r_{\rm m} \propto M_{\rm h}^{-0.21}$, which is obtained from Eq.~\ref{tilde_rc_low_Mhdep}.
We analytically obtain $M_\mathrm{c} \propto M_\mathrm{h}^{0.27}$.
By fitting numerical calculation, we obtain $M_{\rm c} \propto M_{\rm h}^{0.32}$, from which we find that the analytic estimate above is roughly consistent.

In summary, with the Relaxation model, we find that the CHMR in the FDM halos roughly obeys the double power law, 
\begin{eqnarray}
    && M_{\rm c} \propto M_{\rm h}^{0.32}\ \ \ \ \ \ \ \ \text{low-mass halos} \label{Mc_low_Mhdep} \\
    && M_{\rm c} \propto M_{\rm h}^{0.07}\ \ \ \ \ \ \ \ \text{high-mass halos}. \label{Mc_high_Mhdep}
\end{eqnarray}
The turning halo mass corresponds to four times the half-mode mass $M_{\rm h}^{\rm 4hm}$, where the $c_{\rm vir}$-$M_{\rm h}$ shows the turnover.
The power law index of the CHMR also changes around the boundary mass defined in Eq.~\ref{bound_mass}, but this change is more modest than that around $M_{\rm h}^{\rm 4hm}$.
Moreover, the boundary mass is almost the same as $M_{\rm h}^{\rm 4hm}$ at redshift $z=0$, as discussed before.

\subsubsection{Scatter} \label{subsubsec:scatter_analysis}
In the previous section, we use the mean $c_{\rm vir}$-$M_{\rm h}$ relation to obtain the semi-analytic behavior of the CHMR and show that it can be expressed in a double power law.
Here, we consider the scatter of the $c_{\rm vir}$-$M_{\rm h}$ relation and examine how it propagates to the degree of the scatter in the CHMR.
Note again that we assume the scatter of the $c_{\rm vir}$-$M_{\rm h}$ relation is 0.16 dex for 1$\sigma$. 

Since we only change the magnitude of the concentration, the power law index of the $c_{\rm vir}$-$M_{\rm h}$ relation is not changed.
However, the halo mass dependence of $f(c_{\mathrm{vir}})$ is slightly changed, which causes the change of that of the relevant combinations of the NFW parameters as $4\pi\rho_{\rm s}r_{\rm s}^{3}|_{{\rm low} M_{\rm h}} \propto M_{\rm h}^{0.78 + 0.025n}$, and $2\pi\rho_{\rm s}r_{\rm s}|_{{\rm high} M_{\rm h}} \propto M_{\rm h}^{0.16 - 0.01n}$.
While the power law index in the former combination becomes larger as the magnitude of the concentration becomes larger, the opposite behavior can be seen in the latter combination.
These differences in the halo mass dependence would result in the different power law indices of the CHMR.
The detailed analysis to obtain the power law index for each case is totally the same as the mean case as we show in Sec.\ref{subsubsec:mean_analysis}.
We finally obtain the following relation by fitting,
\begin{eqnarray}
    &&M_{\rm c} \propto M_{\rm h}^{0.32 + 0.05n}\ \ \ \ \ \ \ \ \text{low-mass halos}\\
    &&M_{\rm c} \propto M_{\rm h}^{0.07 - 0.01n}\ \ \ \ \ \ \ \ \text{high-mass halos}.
\end{eqnarray}
As expected, the dependence of the power law index on the scatter is the opposite between the low and high-mass halos.  

In Fig.~\ref{fig:relax_z0}, we also find that the three lines, which correspond to the different magnitudes of the concentration, intersect at almost the same halo mass, $M_{\rm h} \simeq 10^{7.5}\ h^{-1}M_{\odot}$.
This halo mass corresponds to the minimum halo mass as we discuss in the following.
We define the minimum halo mass $M_{\rm h}^{\rm min}$ where the matching radius $r_{\rm m}$ equals the virial radius $r_{\rm vir}$. 
For halos with $r_{\rm m} > r_{\rm vir}$ we cannot use $M_{\rm h}$ in Eq.~\ref{rvir_def}.
From the condition $r_{\rm m} = r_{\rm vir}$ and Eq.~\ref{relax_def_re}, we analytically obtain 
\begin{align}
    M_{\rm h}^{\rm min}(z) & \simeq 4 \times 10^{7}\ M_{\odot} (1+z)^{\frac{3}{4}} \left(\frac{\zeta(z)}{\zeta(0)}\right)^{\frac{1}{2}} \nonumber \\ 
    & \quad \times \left(\frac{mc^{2}}{10^{-22}\ {\rm eV}}\right)^{-\frac{3}{2}} \left(\frac{p_{1}}{0.11}\right)^{1.8} \left(\frac{1}{p_{1}^{0.6}} \frac{f(c_{\mathrm{vir}})}{f(\Tilde{c}_{\mathrm{vir}})}\right)^{\frac{1}{2}} \nonumber \\
    & \quad \times \left(\frac{\rho_{\rm m0}}{40.8\ M_{\odot}\ {\rm kpc^{-3}}}\right)^{\frac{1}{2}} \left(\frac{t_{0}}{13.8\ {\rm Gyr}}\right)^{\frac{1}{2}} \label{min_mass},
\end{align}
for a given redshift $z$.
Here we define $\Tilde{c}_{\mathrm{vir}} = p_{1}^{-1}c_{\mathrm{vir}}$.
The scatter of the minimum halo mass according to that of the concentration parameter is rather small since the factor $f(c_{\mathrm{vir}})/f(\Tilde{c}_{\mathrm{vir}})$ does not largely change.
The core mass of the minimum halo is almost the same as the halo mass by the definition and $M_{\rm c}^{\rm min} \simeq 0.34 M_{\rm h}^{\rm min}(n=0)$, more precisely.
This is why the three lines in Fig.~\ref{fig:relax_z0} intersect around the minimum halo mass in the FDM halos.
We check that the mass of the intersection in the numerical calculation coincides with the analytically estimated minimum halo mass with different redshift and FDM mass within a factor of $\mathcal{O}(1)$.
Note again that though we plot the CHMR below the minimum halo mass, our approach may not be valid there.
On the other hand, we do not have any upper limits to the core and halo mass since 
there is no lower limit to the core radius in our model.

Before we end this section, we comment on the scatter of the boundary halo mass Eq.~\ref{bound_mass}.
We find that the boundary mass gets about 5 times larger/smaller than the mean case when considering the $2/-2 \sigma$ scatter.
On the other hand, we find that the mass where the power law index changes in the CHMR does not change significantly.
This is why we conclude in Sec.~\ref{subsubsec:mean_analysis} that the change in the power law index around the boundary mass is more modest than that around four times the half-mode mass $M_{\rm h}^{\rm 4hm}$ where the $c_{\rm vir}$-$M_{\rm h}$ relation shows the turnover.

\begin{figure}
    \includegraphics[width=\columnwidth]{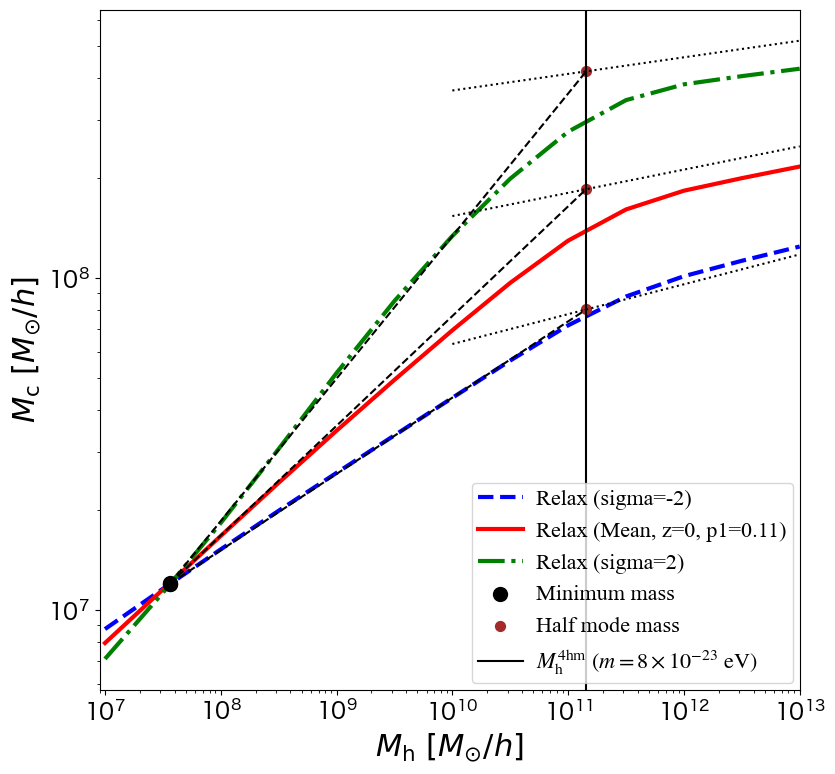}
    \caption{The CHMR in the Relaxation model and its semi-analytic expression at redshift $z=0$.
    The red solid, green dash-dotted, and blue dashed lines are the same as Fig.~\ref{fig:relax_z0}.
    The black big point indicates the minimum halo mass and the brown small points indicate the four times the half-mode mass.
    The black dashed and dotted lines show the analytic expressions in Eq.~\ref{analytic_expression}.
    }
    \label{fig:analytic_model}
\end{figure}

Combining all our results in Secs.~\ref{subsubsec:mean_analysis} and \ref{subsubsec:scatter_analysis}, we can express the CHMR at redshift $z=0$ in a simpler semi-analytic way as;
\begin{equation}
    M_{\rm c} = 
    \begin{cases}
        M_{\rm c}^{\rm min} \left(\frac{M_{\rm h}}{M_{\rm h}^{\rm min}}\right)^{0.32 + 0.05 n}\ (M_{\rm h}^{\rm min} < M_{\rm h} < M_{\rm h}^{\rm 4hm}) \\
        M_{{\rm c}, n}^{\rm 4hm} \left(\frac{M_{\rm h}}{M_{\rm h}^{\rm 4hm}}\right)^{0.07 - 0.01 n} (M_{\rm h} > M_\mathrm{h}^{\rm 4hm})
    \end{cases} \label{analytic_expression}
\end{equation}
where we define $M_{{\rm c}, n}^{\rm 4hm}$ such that this expression is continuous at $M_{\rm h}=M_{\rm h}^{\rm 4hm}$ for a given $n$.
Figure~\ref{fig:analytic_model} indicates that the semi-analytic expression in Eq.~\ref{analytic_expression} reproduces the Relaxation model within the error of $\mathcal{O}(1)$.

\subsubsection{FDM mass} \label{subsubsec:fdm_mass_analysis}
Another important discussion on the CHMR is the dependence of the FDM mass, which is constrained by many observations, such as the rotation curve measurements and the gravitational lensing events.

The FDM mass dependence of the Relaxation model can be understood in the same way as the previous analysis.
Since we already know the FDM mass dependence of the minimum mass and the half-mode mass, where the CHMR shows the transition of the double power law behavior, the only part that we need additional discussion on is how the power law index depends on the FDM mass.  
Note that the relevant combinations of the NFW parameters such as the enclosed mass include the logarithmic dependence on the halo mass, and hence we rely on fitting to obtain the exact value for the power law index. 
Indeed, we find that there is almost no dependence of the FDM mass on the power law index of the CHMR within the range of interest.

\begin{figure}
    \includegraphics[width=\columnwidth]{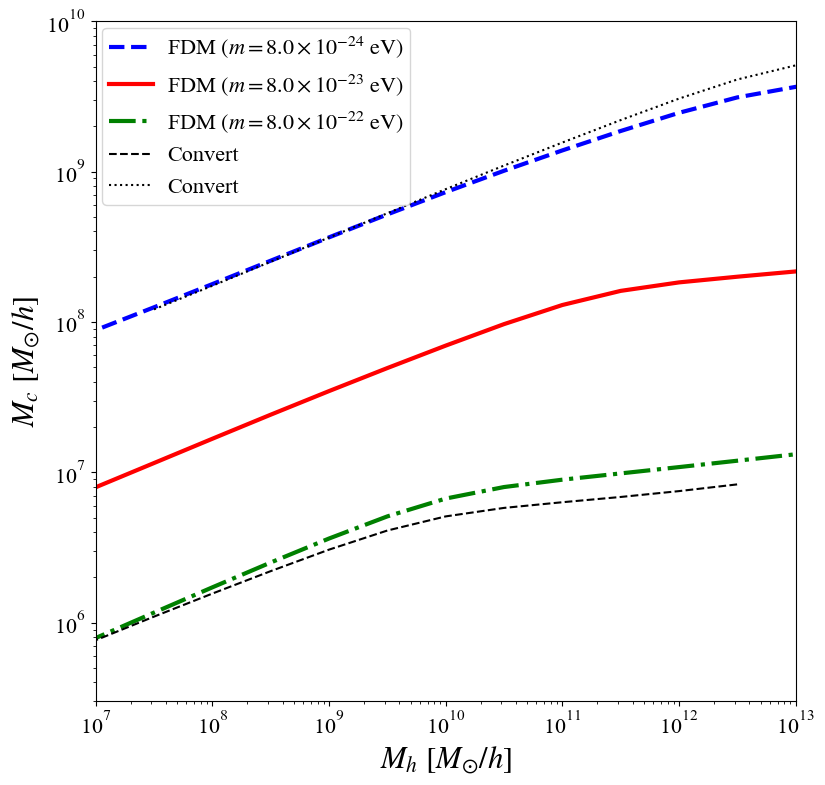}
    \caption{The comparison between the Relaxation model and the scaling relation of the SP equation.
    The blue dashed, red solid, and green dash-dotted lines are the Relaxation models with different FDM masses. 
    The concentration parameter is taken to be the mean value.
    The black dotted and dashed lines are plotted by applying the scaling relation $M \propto m^{-3/2}$ to the red solid line.
    }
    \label{fig:FDMmass_dep}
\end{figure}

The minimum halo and core masses roughly scale as $M^{\rm min} \propto m^{-3/2}$ as can be seen in Eq.~\ref{min_mass}, and the half-mode mass scales as $M_{\rm h}^{\rm 4hm} \propto m^{-4/3}$ as shown in Eq.~\ref{half_mode_mass}.
The core mass corresponding to the halo mass with four times the half-mode mass scales as $M_{{\rm c},n}^{\rm 4hm} \propto m^{-(8.68 - 0.05n)/6}$, which can be obtained from the upper relation in Eq.~\ref{analytic_expression} with $M_{\rm h} = M_{\rm h}^{\rm 4hm}$.
These are roughly consistent with the scaling relation of the SP equation,  $M_{\rm h}\propto m^{-3/2}$ and $M_{\rm c}\propto m^{-3/2}$, which can be obtained by setting $\beta = 1$ in Eq.~\ref{sp_scale_sym}.
This consistency can be seen in Fig.~\ref{fig:FDMmass_dep}, which shows the comparison between the scaling relation and the Relaxation model.
The red solid line is plotted with our fiducial FDM mass. 
The green dash-dotted and blue dashed lines are plotted with ten times larger and smaller than the fiducial mass, respectively.
The black dashed and dotted lines are plotted by applying the scaling relation to the red solid line.
It can be concluded that the scaling features of the SP equation are well captured in the Relaxation model.

To be precise, we see slight deviations between the scaling relation of the SP equation and the Relaxation model.
This deviation originates from the $c_\mathrm{vir}$-$M_{\rm h}$ relation in the FDM halos, Eq.~\ref{cfdm_def}.
Both of the two factors in the right-hand side of the Eq.~\ref{cfdm_def}, $c_\mathrm{vir}(M_{\rm h},z; {\rm CDM})$ and $F(M_{\rm h}/M_{\rm h}^{\rm hm})$, break the scaling relation.
We can check them as follows.
For the first factor, the concentration is unchanged under the scaling, while the halo mass changes.
Since the concentration is not constant over all halo masses, which originates from the scale-variant power spectrum, the scaling relation breaks.
Even though the power spectrum of the primordial density perturbation is scale-invariant, it becomes scale-variant due to the evolution of perturbations at the super-horizon and the evolution in radiation dominant epoch, which are characterized by the (CDM) transfer function.
For the second factor, the half-mode mass defined in Eq.~\ref{half_mode_mass} does not respect the scaling relation, which originates from the violation of the scaling law in the FDM transfer function. 
It is modified by $T_{\rm F}(k)$ compared to the CDM transfer function as we can expect from Eq.~\ref{transf}. 
This modification factor incorporates the factor of $(mc^2/10^{-22}\ \mathrm{eV})^{1/18}$, which breaks the scaling relation for the same reason as for the first factor.
Therefore, the Relaxation model deviates from the scaling relation in the SP equation due to these two factors.
Note that in the semi-analytic expression of the Relaxation model, Eq.~\ref{analytic_expression}, the former contribution is found to be negligible in the FDM mass range of our interest, otherwise, the power law indices of the CHMR are expected to be FDM mass-dependent.

The previous simulations are conducted with almost the same FDM mass, our fiducial mass, and with the CDM initial condition, which incorporates only the first factor.  
Therefore, it is important to conduct simulations with different FDM masses and with the FDM initial condition, from which we expect to test the FDM mass dependence of the CHMR.

\subsubsection{Redshift} \label{subsubsec:redshift_analysis}
So far we mainly focus on the CHMR at redshift $z=0$.
In this subsection, we briefly discuss the redshift dependence of the Relaxation model.

In the Relaxation model, the redshift dependence exists in the concentration parameter, the halo age, and $p_1$.
The concentration parameter scales as $c_{\rm vir} \propto (1+z)^{-1}$ and the power law index of the $c_{\rm vir}$-$M_{\rm h}$ does not change for the low-mass halos.
On the other hand, the redshift dependence of the $c_{\rm vir}$-$M_{\rm h}$ relation is not monotonic for higher-mass halos, since such high halos are rare at a higher redshift.
Nevertheless, we can understand the behavior in the same way as the previous analyses. 
In this subsection, we simply show the consequence of the Relaxation model without explaining the details of the analysis.

\begin{figure}
    \includegraphics[width=\columnwidth]{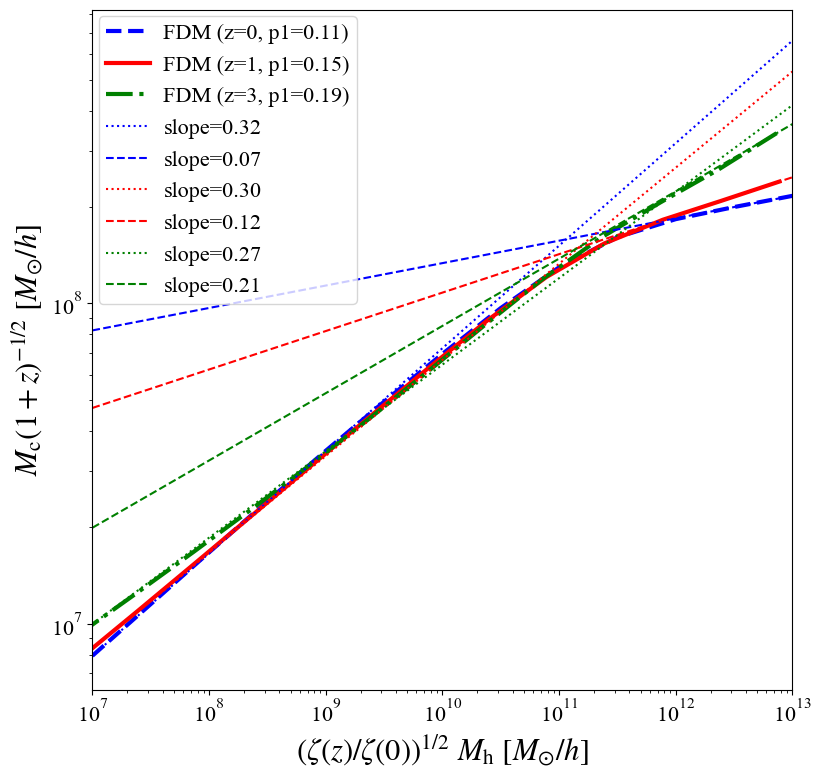}
    \caption{The redshift dependence of the CHMR with different $p_{1}$.
    We show the results of $z=0$ (blue dashed), $z=1$ (red solid), and $z=3$ (green dash-dotted).
    We set $p_{1} = 0.11$ at $z=0$, $0.15$ at $z=1$, and $0.19$ at $z=3$ to absorb the redshift dependence.
    }
    \label{fig:redshift_dep2}
\end{figure}

First, let us fix $p_1 = 0.11$ for simplicity.
\citet{2014PhRvL.113z1302S} states that the redshift dependence of the CHMR can be absorbed by plotting it in the $\sqrt{\zeta(z)/\zeta(0)} M_{\rm h}$-$M_{\rm c}/\sqrt{1+z}$ plane. 
By plotting the result of the Relaxation model in that plane, however, we still see the redshift dependence ($M_{\rm c}/\sqrt{1+z}$ is smaller for a larger redshift).
In addition, the Relaxation model can explain the simulation data obtained by \citet{2021MNRAS.506.2603M}, with $p_{1} = 0.19$ (not 0.11) at $z=3$.
Thus, we allow the redshift dependence of $p_1$.
Since we need a larger core mass at a larger redshift to match three lines, $p_{1}$ must become larger as the redshift becomes larger.
We find that we can absorb the redshift dependence of the CHMR as shown in Fig.~\ref{fig:redshift_dep2}, where we set $p_{1} = 0.11$ at $z=0$, $0.15$ at $z=1$, and $0.19$ at $z=3$.
The resultant power law index of the CHMR in the $\sqrt{\zeta(z)/\zeta(0)} M_{\rm h}$-$M_{\rm c}/\sqrt{1+z}$ plane at low-mass region is about 0.30, which is also roughly consistent with the relation obtained in \citet{2014PhRvL.113z1302S}.

The empirical and practical reason why larger $p_{1}$ is needed is to compensate for the smaller core mass with a shorter halo age $t_{\rm age}$ at a higher redshift.
As we mentioned, the Relaxation model contains the redshift dependence in concentration $c_{\rm vir}$, halo age $t_{\rm age}$, and $p_{1}$.
When $t_{\rm age}(z)$ is fixed to $t_{\rm age}(0)$ in the Relaxation model, we find that the CHMR relations at different redshifts coincide with each other in the $\sqrt{\zeta(z)/\zeta(0)} M_{\rm h}$-$M_{\rm c}/\sqrt{1+z}$ plane.
However, when considering the redshift dependence of $t_{\rm age}$, the core mass becomes smaller at a higher redshift.
Since $p_{1}$ determines the enclosed mass to become the core, we can reproduce the higher core mass by increasing $p_{1}$.

\begin{figure}
    \includegraphics[width=\columnwidth]{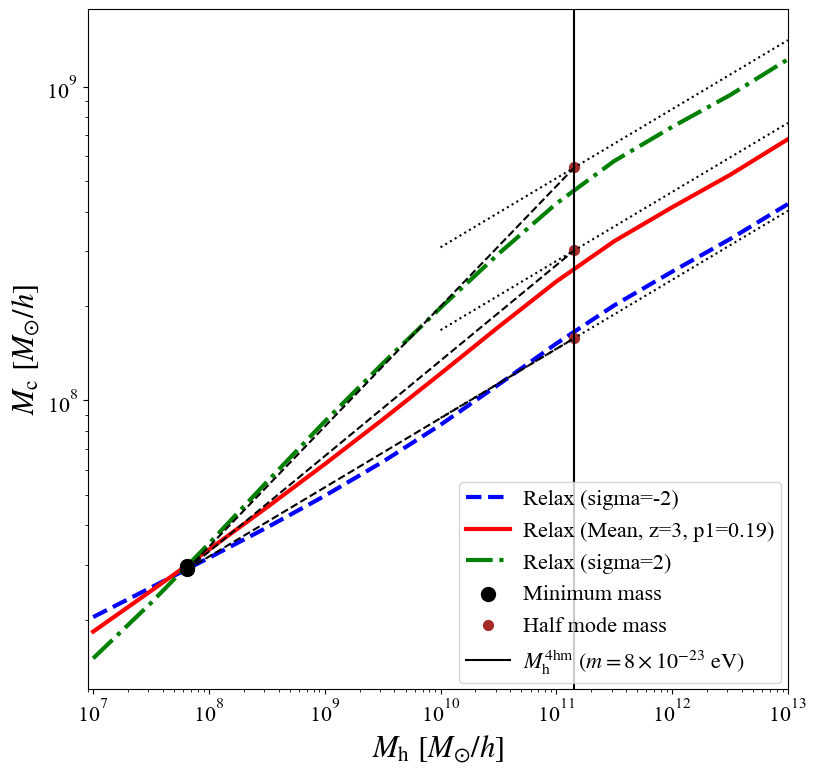}
    \caption{The same as Fig.~\ref{fig:analytic_model}, but in the case of the redshift $z=3$.
    The semi-analytic expression is Eq.~\ref{analytic_expression_z3}, which is plotted by the black dashed and dotted lines.
    Here we show the minimum mass denoted by $\Tilde{M}_{\rm h}^{\rm min}$.
    }
    \label{fig:analytic_model_z3}
\end{figure}

We can obtain the semi-analytic expression at redshift $z=3$ as the same way as in Sec.~\ref{subsubsec:mean_analysis} to Sec.~\ref{subsubsec:scatter_analysis}, 
\begin{equation}
    M_{\rm c} = 
    \begin{cases}
        \Tilde{M}_{\rm c}^{\rm min} \left(\frac{M_{\rm h}}{\Tilde{M}_{\rm h}^{\rm min}}\right)^{0.30 + 0.04 n}\ \  (M_{\rm h}^{\rm min} < M_{\rm h} < M_{\rm h}^{\rm 4hm}) \\
        M_{{\rm c},n}^{\rm 4hm} \left(\frac{M_{\rm h}}{M_{\rm h}^{\rm 4hm}}\right)^{0.22}\ \ \ \ \ \ \ \ \ (M_{\rm h} > M_\mathrm{h}^{\rm 4hm}).
    \end{cases} \label{analytic_expression_z3}
\end{equation}
Here we introduce $\Tilde{M}_{\rm h}^{\rm min} = M_{\rm h}^{\rm min}/2.5$ to fit better to the numerical calculation.
Again the minimum halo mass is estimated with the mean concentration ($n=0$) and we find $\Tilde{M}_{\rm c}^{\rm min} \simeq 0.45\Tilde{M}_{\rm h}^{\rm min}$.
As the case with $z=0$, we define $M_{{\rm c},n}^{\rm 4hm}$ such that this expression is continuous at $M_{\rm h}=M_{\rm h}^{\rm 4hm}$ for a given $n$.
The difference of the power law indices between at redshift $z=0$ (Eq.~\ref{analytic_expression}) and at redshift $z=3$ (Eq.~\ref{analytic_expression_z3}) originates from the different halo mass dependence of the $c_{\rm vir}, f(c_{\rm vir})$ and different $p_{1}$.
While the power law index is almost the same for low-mass halos, that for high-mass halos is largely different.
This is mainly due to the large difference in the halo mass dependence of the concentration.
Since the high-mass halos are rare at a high redshift, the $c_{\rm vir}$-$M_{\rm h}$ relation shows an upturn with increasing halo mass as modeled in \citet{2021MNRAS.506.4210I}, from which we obtain the different power law index.
We also find that the power law index of the high-mass halos is independent of $n$. Note that the scatter of the core mass is incorporated in $M_{{\rm c},n}^{\rm 4hm}$.
In Fig.~\ref{fig:analytic_model_z3} we compare the semi-analytic expression with numerical calculation at redshift $z=3$, showing a good agreement between them.
We use Eq.~\ref{analytic_expression_z3} in Sec.~\ref{subsubsec:compare_other_models} to compare the scatter of the CHMR.

\subsection{Comparison with other (semi-)analytic models} \label{subsubsec:compare_other_models}
In this subsection, we compare the Relaxation model with previous studies and discuss the scatter of the CHMR.

In Fig.~\ref{fig:relax_z0}, we plot the empirical relation obtained in \citet{2014PhRvL.113z1302S}, finding that the power law index of our model for the low-mass halos at redshift $z=0$ is consistent with it.
Moreover, the Relaxation model with different redshift is also consistent with them by allowing the redshift dependence of $p_{1}$ as we mention in the previous subsection.
However, there are discrepancies between the Relaxation model and their result for the high-mass halos. 
The double power law nature is also claimed in \citet{2022PhRvD.106j3532T} with a different scenario; they calculate the linearized SP equation by assuming the NFW potential as a background.
It claims that the power law index is roughly 1/3 in low-mass regions and becomes smaller in high-mass halos, which shows the same behavior as the Relaxation model.
By obtaining more data on the high-mass halos, we might be able to test the models, ours and \citet{2022PhRvD.106j3532T}.

In Fig.~\ref{fig:relax_z3}, we plot the expected scatter of the CHMR empirically obtained by \citet{2022MNRAS.511..943C} in grey dotted lines.
Here you can see that the degree of the scatter is much smaller around the minimum halo mass,  which is well described in the Relaxation model.
Although this trend is also mentioned in \citet{2022PhRvD.106j3532T}, which also considers the impact of the scatter in the $c_{\rm vir}$-$M_{\rm h}$ relation,
it seems that the degree of the changes of the scatter with respect to halo mass is not enough.

To quantitatively compare the degree of the scatter of the CHMR among our model and the simulation results, we consider the distribution of the core mass for a given halo mass.
First, we can derive the distribution and scatter of the CHMR in the Relaxation model as follows.
From the fact that the concentration follows the log-normal distribution with the constant scatter of about $0.16$ dex, the probability distribution of the concentration can be expressed as
\begin{equation}
    dP = \frac{d\log_{10}c_{\rm vir}}{\sqrt{2\pi \sigma_{\log_{10}c_{\rm vir}}^{2}}} \exp \left(-\frac{(\log_{10}c_{\rm vir}-\log_{10}c_{\rm vir}^{\rm mean})^{2}}{2\sigma_{\log_{10}c_{\rm vir}}^{2}}\right) \label{log-normal_c},
\end{equation}
where $\sigma_{\log_{10}c_{\rm vir}}=0.16$ is the variance of the concentration.
Since we use the relation $c_\mathrm{vir} = c_\mathrm{vir}^{\rm mean} \times 10^{0.16 n}$, we can express as 
\begin{equation}
    n = \sigma_{\log_{10}c_{\rm vir}}^{-1} \log_{10}\left(\frac{c_{\rm vir}}{c_{\rm vir}^{\rm mean}}\right) \label{log-normal_scatter_n}.
\end{equation}
Using the semi-analytic expression Eqs.~\ref{analytic_expression_z3}, \ref{log-normal_c} and \ref{log-normal_scatter_n}, we can show that the core mass follows the log-normal distribution,
\begin{equation}
    dP = \frac{d\log_{10}M_{\rm c}}{\sqrt{2\pi \sigma_{\log_{10}M_{\rm c}}^{2}}} \exp \left(-\frac{(\log_{10}M_{\rm c}-\log_{10}M_{\rm c}^{\rm mean})^{2}}{2\sigma_{\log_{10}M_{\rm c}}^{2}}\right),
\end{equation}
where $M_{\rm c}^{\rm mean}$ denotes the core mass calculated with the mean concentration, and the scatter is expressed as 
\begin{equation}
    \sigma_{\log_{10}M_{\rm c}} =
    \begin{cases}
        0.04 \log_{10} \left(\frac{M_{\rm h}}{\Tilde{M}_{\rm h}^{\rm min}}\right) \ \  (M_{\rm h}^{\rm min} < M_{\rm h} < M_{\rm h}^{\rm 4hm}) \\
        0.04 \log_{10} \left(\frac{M_{\rm h}^{\rm 4hm}}{\Tilde{M}_{\rm h}^{\rm min}}\right)\ \ (M_{\rm h} > M_\mathrm{h}^{\rm 4hm}).
    \end{cases} \label{analytic_expression_scatter}
\end{equation}
From Eq.~\ref{analytic_expression_scatter}, the scatter of the core mass becomes larger as the halo mass becomes larger in low-mass halos and is constant in high-mass halos as we expected.
Note that since the dependence of $c_{\rm vir}$-$M_{\rm h}$ relation on the CHMR is monotonic, the core mass calculated for the $n \sigma$ variance of the concentration corresponds to the same variance of the core mass as we plot in Figs.~\ref{fig:jeans_z0} to \ref{fig:relax_z3}.
Hereafter we take $z=3$, namely, Eq.~\ref{analytic_expression_z3}, to compare the Relaxation model with the simulation results \citep{2021MNRAS.506.2603M}.
On the other hand, the same analysis with Eq.~\ref{analytic_expression} gives the log-normal distribution also for $z=0$, but with a different expression for the scatter.

\begin{figure}
    \includegraphics[width=\columnwidth]{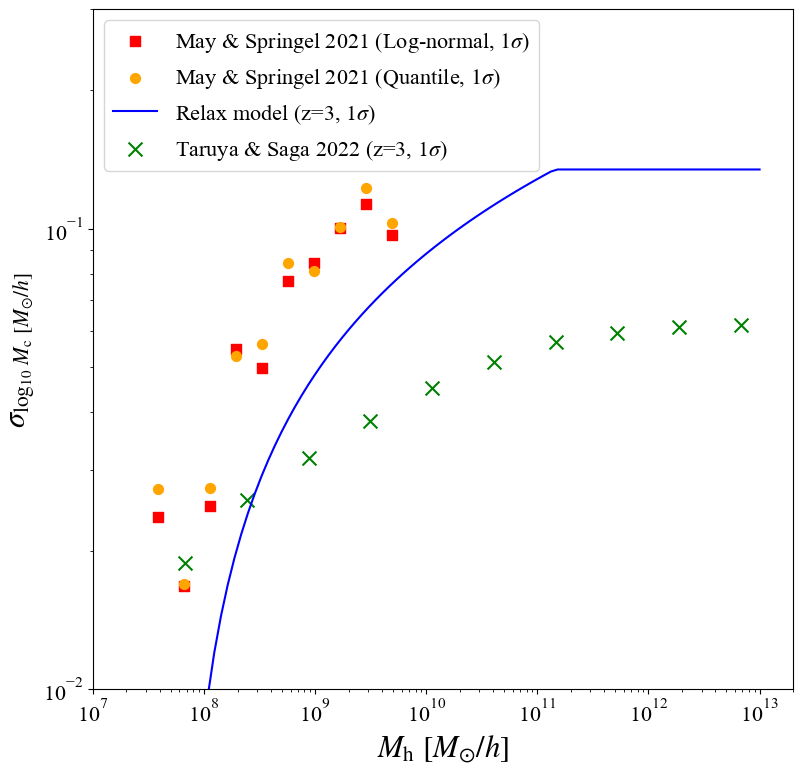}
    \caption{The degree of $1\sigma$ scatters of the core mass at redshift $z=3$ is compared.
    The red square and orange circle dots represent the simulation by \citet{2021MNRAS.506.2603M}, where the former is estimated by fitting the log-normal distribution, and the latter is estimated by computing empirical quantiles of the 68$\%$ around the median core mass.
    The blue solid line is the prediction of the Relaxation model obtained by Eq.~\ref{analytic_expression_scatter}.
    The green dots with the cross symbol are estimated by \citet{2022PhRvD.106j3532T}.
    Although both the Relaxation model and the model by \citet{2022PhRvD.106j3532T} cannot fully explain the scatter of the CHMR, the Relaxation model shows a similar trend to the simulation data.
    }
    \label{fig:Mc_variance}
\end{figure}

We find that the analytic model shown in \citet{2022PhRvD.106j3532T} also follows the log-normal distribution as shown in Fig.~\ref{fig:Taruya_Mc_pdf}.
Moreover, we find that the distribution of the core mass in the simulation data \citep{2021MNRAS.506.2603M} follows the log-normal distribution as shown in Fig.~\ref{fig:May_Mc_pdf}, which might indicate that the Relaxation model and the model by \citet{2022PhRvD.106j3532T} capture the features well. 
We compare the $1 \sigma$ scatter of the core mass obtained by those studies and the Relaxation model as shown in Fig.~\ref{fig:Mc_variance}. 
The blue line shows the variance of the core mass predicted by the Relaxation model in Eq.~\ref{analytic_expression_scatter}.
As we mentioned before, the scatter of core mass increases with increasing halo mass for low-mass halos and remains constant for high-mass halos.
The green points dotted with a cross show the prediction obtained by \citet{2022PhRvD.106j3532T}, which also shows the same trend as the Relaxation model but the amplitude of the variance is smaller than that of the Relaxation model except for the halo mass below $\sim 10^{8}\ M_{\odot}/h$.
We take two approaches to estimate the scatter of core mass for a given halo mass in the simulation data.
One is fitting the distribution with the log-normal function to obtain the scatter, which is shown in red square dots.
The other is computing quantiles for the data, which is the same procedure conducted by \citet{2022PhRvD.106j3532T}.
We identify two quantiles: 16\% of the data is contained below one and above the other.
We take the half of the difference of two quantiles as 1$\sigma$.
We show the result by the orange circle dots.
Since the distribution of the core mass is in reasonable agreement with the log-normal distribution, these two approaches to estimating the scatter yield almost identical results.
In Fig.~\ref{fig:Mc_variance}, we find that both the Relaxation model and the model by \citet{2022PhRvD.106j3532T} with the scatter of the $c_{\rm vir}$-$M_{\rm h}$ relation do not fully explain the scatter of the CHMR obtained by the simulation.
While the scatter of the core mass in the simulation is 1.5 times larger than the Relaxation model, the dependence of the scatter on the halo mass seems in good agreement between them.
Therefore, we can conclude that the scatter of the $c_{\rm vir}$-$M_{\rm h}$ relation partially contributes to the scatter of the CHMR.
Indeed, other effects, such as the shape of the halo, are also thought to contribute to the scatter of the core mass in the simulations.
Since the Relaxation model has only one degree of freedom for the scatter, the concentration, we may need to extend the model, which we leave for future study.

\section{Summary and discussions} \label{sec:summary_disscusion}
We have developed a semi-analytic model of the CHMR in the FDM halos.
We assume that the soliton core forms through mass redistribution in the FDM halo whose initial density distribution is given by the NFW profile.
Our model first assumes the empirical $c_{\rm vir}$-$M_{\rm h}$ relation to fix the NFW profile. 
We adopt the model of \citet{2022MNRAS.515.5646D} that incorporates the relative reduction of the concentration at small masses compared to the CDM halos.
We then calculate the characteristic radius by considering two physically motivated conditions, which we dub the Jeans model and the Relaxation model.
The matching radius within which the mass redistribution occurs is defined as $p_{1}$ times the characteristic radius, and $p_{1}$ is assumed to be constant for all halos but is allowed to depend on the redshift.
We then obtain the mass of the soliton core within the matching radius, which is equal to that of the original NFW profile.
The Relaxation model reproduces the results of the recent cosmological simulations by \citet{2014NatPh..10..496S} and \citet{2021MNRAS.506.2603M} if we set $p_{1} = 0.11$ and $0.19$ at redshift $z=0$ and $z=3$, respectively. 
We also consider the scatter of the $c_{\rm vir}$-$M_{\rm h}$ relation in our models. 
The degree of the scatter in the Relaxation model is larger than that in the Jeans model.

The CHMR in the Relaxation model follows roughly a double power law changing the index at four times the half-mode mass of FDM.
If we use the mean $c_{\rm vir}$-$M_{\rm h}$ relation, the CHMR is consistent with that of \citet{2014NatPh..10..496S} for the low-mass halos, but the power law index is smaller for the high-mass halos.
The difference of the power law indices originates from whether the characteristic radius is bigger than the scale radius of the original NFW profile, and from the the turnover nature of the $c_{\rm vir}$-$M_{\rm h}$ relation.
We find that the latter affects more strongly the shape of the CHMR.
The CHMRs with different magnitudes of the concentration parameter intersect around a single mass scale corresponding to the minimum halo mass, where the matching radius equals the halo's virial radius.
The Relaxation model also reproduces the sizable scatter in the CHMR by considering the scatter of the $c_{\rm vir}$-$M_{\rm h}$ relation. 
The scatter of the CHMR is typically smaller at around the minimum halo mass.
The resulting semi-analytic expressions are given by Eqs.~\ref{analytic_expression} and \ref{analytic_expression_z3} at redshift $z=0$ and $z=3$, respectively.
We find that the core mass follows the log-normal distribution, both in the Relaxation model (where we can derive it from the semi-analytic formula) and in simulation results.
Although the degree of the scatter is not enough compared to the simulation data, the Relaxation model can well reproduce the trend of the scatter of the CHMR with varying halo mass.
We thus argue that the scatter of the core mass at a given halo mass found in the previous numerical simulations may partially originate from the scatter of the concentration of the halo, which itself likely originates from the assembly history of individual halos.
There may be other effects, such as the shape of halos, besides the scatter of the concentration as the reason for core mass to distribute in the log-normal function and to have larger scatter in the simulation. 
Since the degree of freedom for the scatter in our model is only the concentration, we may need to extend the model, $\it{e.g.}$, non-spherical halos, which we leave for future study.

We have found that the best-fit value for $p_{1}$ is $0.11$ in the Relaxation model at redshift $z=0$.
This value is reasonable since many granular structures should exist inside the characteristic radius and $p_{1}$ should be below $1$.
Even though it successfully reproduces the expected results, the resultant small value of $p_{1}$ may indicate that the estimate of the relaxation time can be better formulated.
In the Jeans model, the best-fit value of $p_{1}$ is $2.35$ at redshift $z=0$. 
Since the characteristic radius in the Jeans model corresponds to the de Broglie wavelength, which is around but smaller than the core/mass-matching radius, $p_{1}$ should be around $2$-$3$.
The value of $p_{1} = {\cal O} (1)$ may indicate that the hydrostatic equilibrium condition is physically well-motivated to determine the characteristic/core radius as discussed in the literature.
However, the Jeans model does not provide a sizable scatter of the CHMR under the assumption of mass redistribution.
It would be interesting to develop modified/additional conditions so that the Jeans model can reproduce the CHMR found in numerical simulations. 
We find that $p_{1}$ becomes larger as the redshift becomes larger in the Relaxation model.
The empirical and practical reason for this behavior is to compensate for the smaller core mass with a shorter halo age.
Here we question the validity of $c_{\rm vir}$-$M_{\rm h}$ relation in FDM halos, where we apply the suppression below the half-mode mass as the model by \citet{2022MNRAS.515.5646D}.
With this model, the concentration parameter becomes below 1 in low-mass halos at a higher redshift, which is physically unacceptable.
For the same reason, we decide not to refer to \citet{2022MNRAS.517.1867L} for $c_{\rm vir}$-$M_{\rm h}$ relation in FDM halos.
We expect that those halos, in reality, have a larger concentration, $c_{\rm vir} > 1$.
Therefore, we may underestimate the concentration at a higher redshift.
Since the core mass becomes smaller with smaller concentration, we might need to tune it by choosing larger $p_{1}$.
To understand the redshift dependence of $p_{1}$ more in detail, we need to study the redshift dependence of the $c_{\rm vir}$-$M_{\rm h}$ relation in FDM halos from the simulation and/or semi-analytic approach.
Thus, we leave it for future work.

We expect that our model can be applied to more realistic cases, for example, with additional contributions from baryons.
To this end, it would be useful to use the results of a FDM simulation which includes baryonic physics \citep{2020PhRvD.101h3518V}.
A promising application of our model may be fitting the rotation curves of galaxies.
Several studies already attempted to use galaxy rotation curves to constrain the FDM mass ({\it{e.g.,}} \citet{2018MNRAS.475.1447B, 2018PhRvD..98h3027B, 2019PhRvD..99j3020B, 2021ApJ...912L...3H, 2022PhRvD.105h3015B, 2023MNRAS.523.3393K}).
It would be important to combine the description of the scatter of the CHMR given in this study and the impact of the baryon potential, to test the validity of the FDM model more preciously. 
Strong gravitational lensing may be another interesting application.
Because of the characteristic core structure, 
FDM would imprint different lensing signatures.
Related studies already exist in the literature ({\it{e.g.,}} \citet{2022MNRAS.517.1867L, 2023NatAs...7..736A}).
Since the FDM soliton core radius becomes smaller as the halo mass becomes larger, it is expected that the effect of the massive soliton core in a galaxy cluster on the strong lens may be relatively small.
However, we have shown that the CHMR is shallower for massive, cluster-size halos, indicating that the core radius may be larger than previously expected.
It would be interesting to explore the possibility of detecting imprints of FDM on the galaxy cluster lens images.

\begin{acknowledgments}
We thank the anonymous referee for the useful comments.
We thank Hei Yin Jowett Chan and Simon May for providing us with their simulation samples of the FDM halos and for discussing the CHMR.
We thank Masamune Oguri, Tilman Hartwig, and Atsushi Taruya for giving us useful feedback during this project.
H. K. is supported by JSPS KAKENHI Grant Number JP22J21440.
H. K. thanks the hospitality of INAF OAS Bologna where part of this work was carried out.
It is supported by JSPS KAKENHI Grant Numbers JP22K21349.
A. K. acknowledges partial support from Norwegian Financial Mechanism for years 2014-2021, grant nr 2019/34/H/ST2/00707; and from National Science Centre, Poland, grant 2017/26/E/ST2/00135 and DEC-2018/31/B/ST2/02283.
K. K. was supported by the National Natural Science Foundation of China (NSFC) under Grant No. 12347103 and JSPS KAKENHI Grant-in-Aid for Challenging Research (Exploratory) JP23K17687.
\end{acknowledgments}

\bibliography{ref}

\appendix
\section{Core-halo mass relation with the concentration in CDM halos} \label{CHMR_cdm}
\begin{figure}
    \includegraphics[width=\columnwidth]{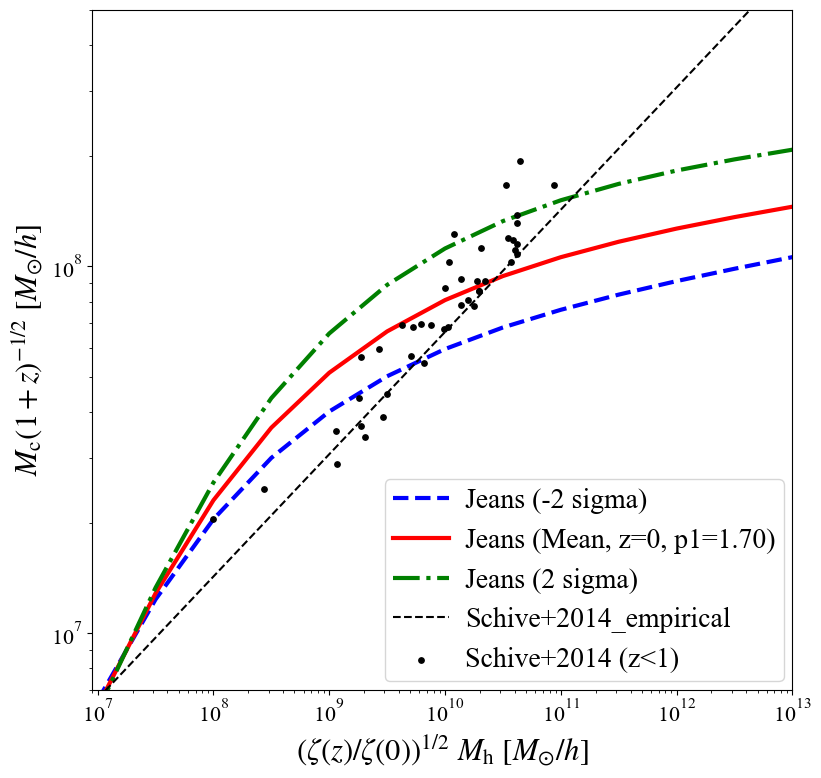}
    \caption{Similar to Fig.~\ref{fig:jeans_z0}, but with the use of $c_{\rm vir}$-$M_{\rm h}$ relation in CDM halos.
    Here we set $p_{1}=1.70$.
    }
    \label{fig:jeans_z0_cdm}
\end{figure}

\begin{figure}
    \includegraphics[width=\columnwidth]{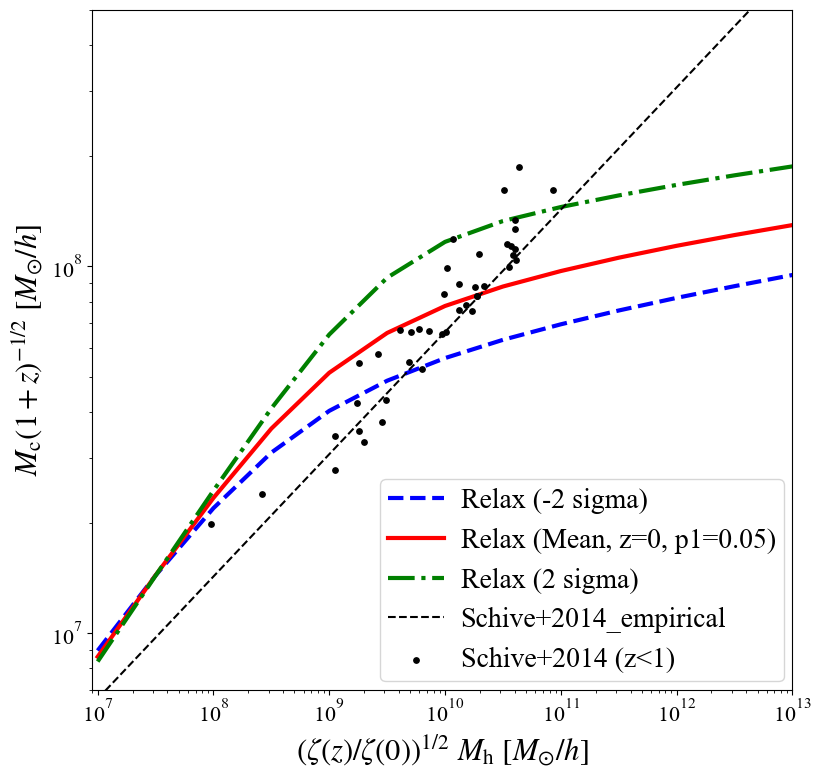}
    \caption{Similar to Fig.~\ref{fig:relax_z0}, but with the use of $c_{\rm vir}$-$M_{\rm h}$ relation in CDM halos.
    Here we set $p_{1}=0.05$. 
    }
    \label{fig:relax_z0_cdm}
\end{figure}

\begin{figure}
    \includegraphics[width=\columnwidth]{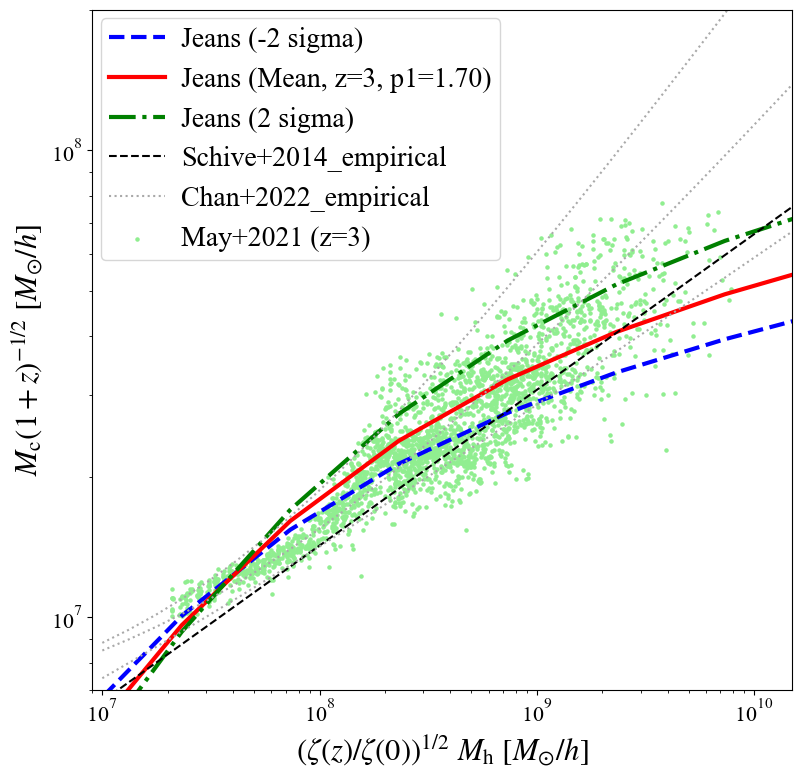}
    \caption{Similar to Fig.~\ref{fig:jeans_z3}, but with the use of $c_{\rm vir}$-$M_{\rm h}$ relation in CDM halos.
    Here we set $p_{1}=1.70$. 
    }
    \label{fig:jeans_z3_cdm}
\end{figure}

\begin{figure}
    \includegraphics[width=\columnwidth]{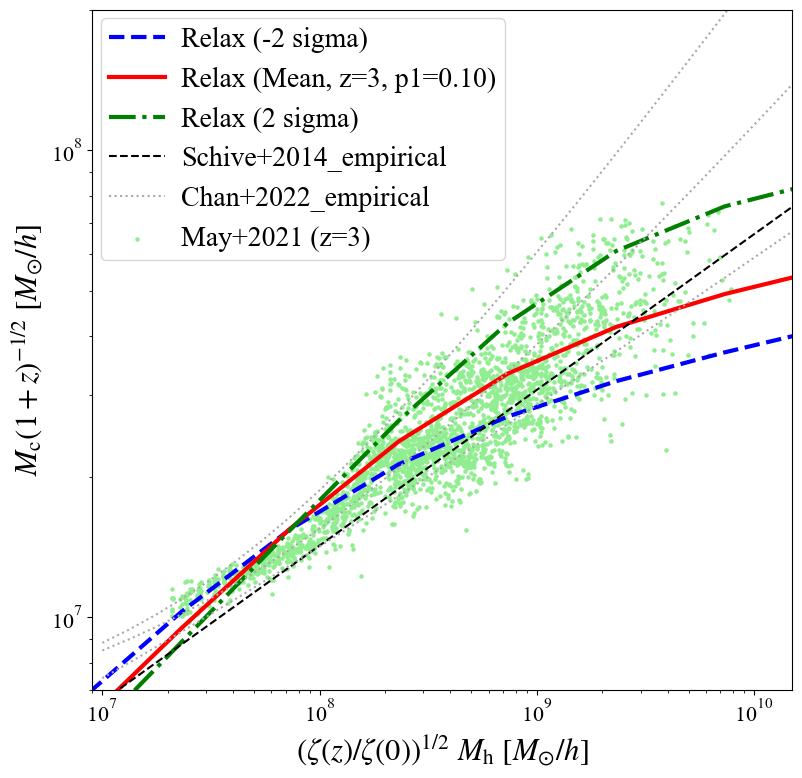}
    \caption{Similar to Fig.~\ref{fig:relax_z3}, but with the use of $c_{\rm vir}$-$M_{\rm h}$ relation in CDM halos.
    Here we set $p_{1}=0.10$. 
    }
    \label{fig:relax_z3_cdm}
\end{figure}

In this section, we show the CHMRs with CDM $c_{\rm vir}$-$M_{\rm h}$ relation obtained by \citet{2021MNRAS.506.4210I}. 
Figures~\ref{fig:jeans_z0_cdm} and \ref{fig:relax_z0_cdm} compare the simulation data obtained by \citet{2014NatPh..10..496S} at redshift $z<1$ and the Jeans/Relaxation model at redshft $z=0$, respectively.
Figures~\ref{fig:jeans_z3_cdm} and \ref{fig:relax_z3_cdm} compare the simulation data obtained by \citet{2021MNRAS.506.2603M} and the Jeans/Relaxation model at redshift $z=3$, respectively.
The mean CHMR in the Jeans/Relaxation model are in reasonable agreement with the simulation data at $z=3$ by setting $p_{1}=1.70$ and $0.05$, respectively.
On the other hand, we could not find values of $p_{1}$ with which the agreement between the model mean CHMR and the simulation data is as good as with FDM $c_{\rm vir}$-$M_{\rm h}$ relation, at redshift $z<1$.

\section{Log-normal distribution}
\begin{figure}
    \includegraphics[width=\columnwidth]{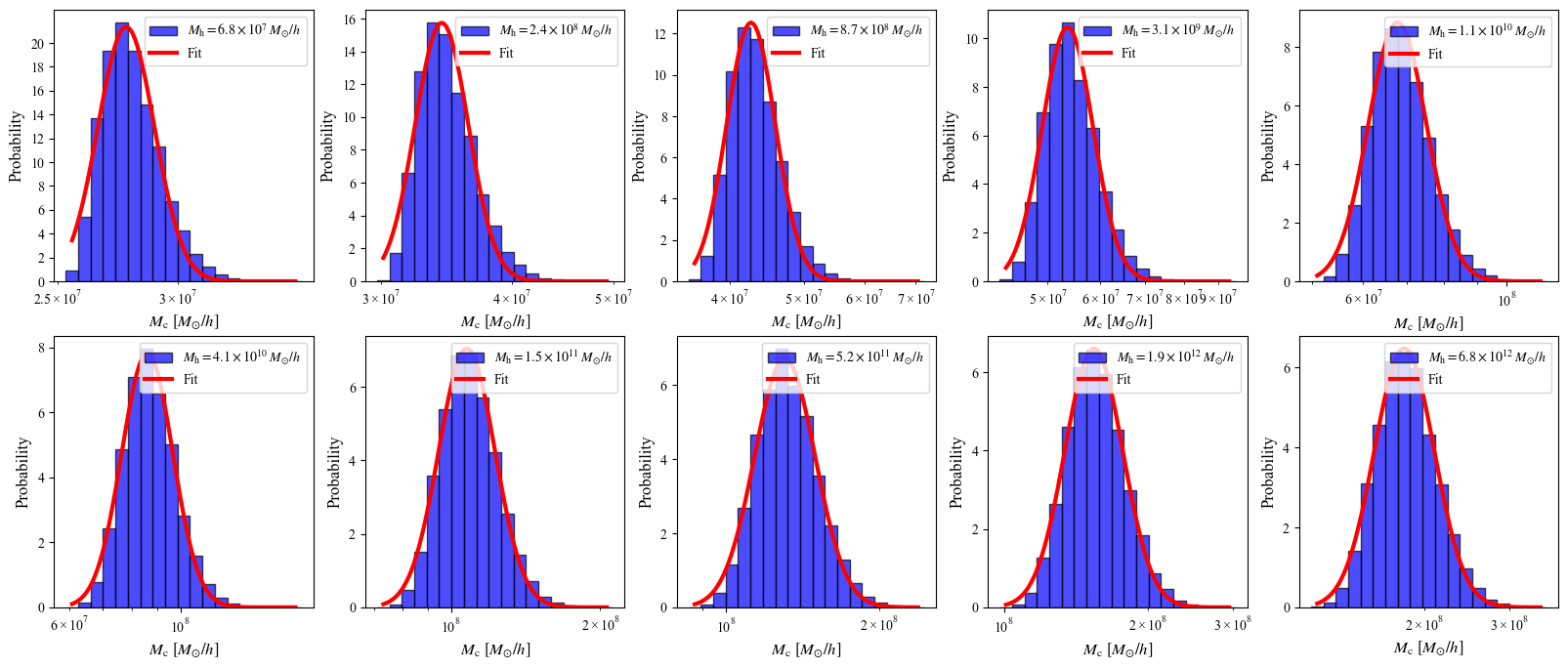}
    \caption{The probability distribution of the core mass with 10 different halo mass bins obtained by the analytic study \citet{2022PhRvD.106j3532T} at redshift $z=3$ by using their public code.
    For a given halo mass, we compute the 100,000 realizations of the core mass by considering the scatter of the concentration parameter.
    The concentration parameters are assumed to follow the log-normal distribution whose mean is obtained by the $c_{\rm vir }$-$M_{\rm h}$ relation \citep{2021MNRAS.506.4210I} and the suppression factor by \citet{2022MNRAS.515.5646D} and variance is 0.16 dex.
    The red lines show the result of the log-normal fitting. 
    }
    \label{fig:Taruya_Mc_pdf}
\end{figure}

\begin{figure}
    \includegraphics[width=\columnwidth]{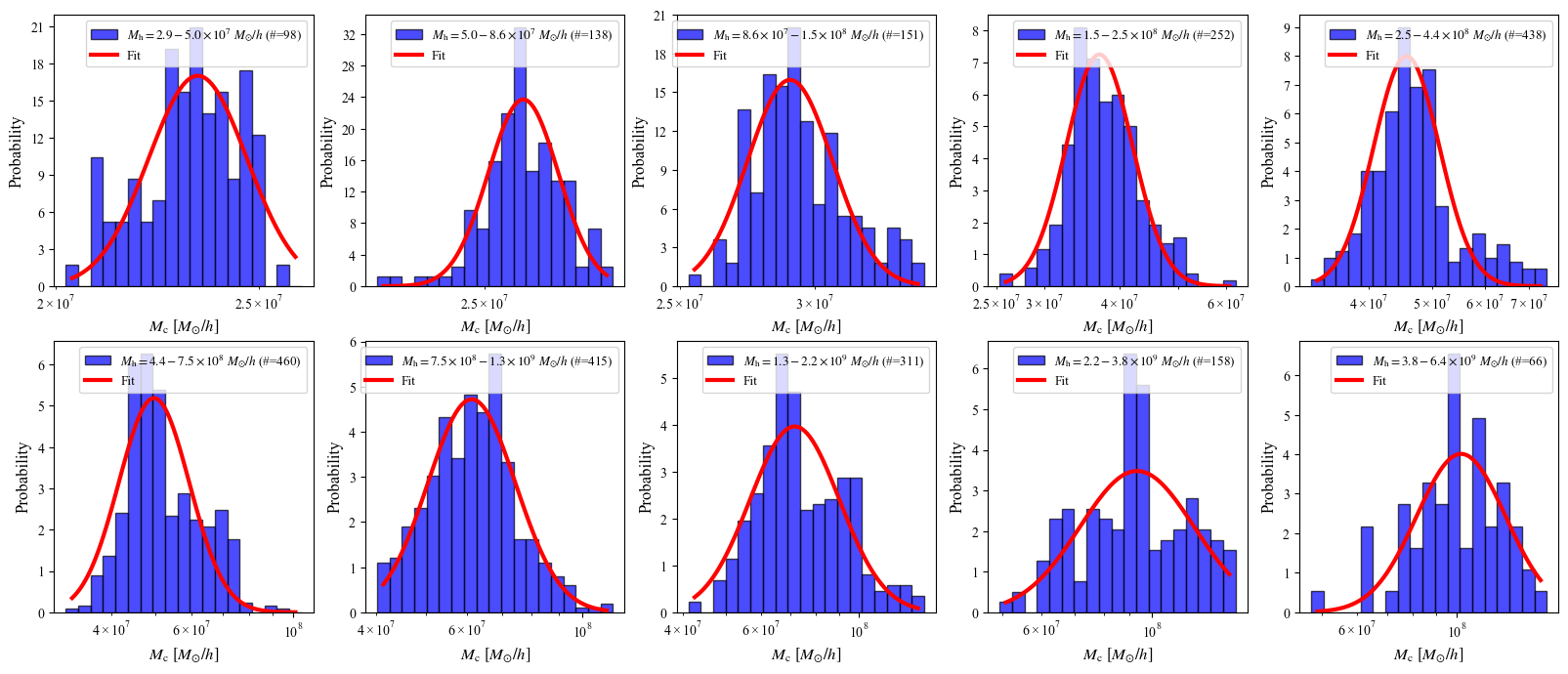}
    \caption{The probability distribution of the core mass with 10 different halo mass bins obtained by the largest FDM simulation, \citet{2021MNRAS.506.2603M}, at redshift $z=3$. 
    The red lines show the result of the log-normal fitting.
    }
    \label{fig:May_Mc_pdf}
\end{figure}

As shown in Sec.~\ref{subsubsec:compare_other_models}, the Relaxation model with the scatter of $c_{\rm vir}$-$M_{\rm h}$ relation predicts that core mass for a given halo mass follows the log-normal distribution function.
To see the agreement between the semi-analytic models and the simulation, we also study the distribution of the core mass obtained analytically in \citet{2022PhRvD.106j3532T} by using their public code, cite https://github.com/ataruya/FDM.
For a given halo mass, we first obtain the 100,000 realizations of the concentration. 
Here we assume that the concentration follows the log-normal distribution whose mean value is determined by the $c_{\rm vir}$-$M_{\rm h}$ relation obtained by the \citet{2021MNRAS.506.4210I} and the suppression factor by \citet{2022MNRAS.515.5646D} and the scatter is 0.16 dex, namely, the same as in ours.
Then we calculate the core masses for each concentration and obtain 100,000 realizations of the core mass. 
We obtain the core mass distribution with 10 different halo masses as shown in Fig.~\ref{fig:Taruya_Mc_pdf}.
We find that the distribution of the core mass is well-fitted by the log-normal distribution function as in the Relaxation model.
In addition, we study the distribution of the core mass in the simulation data by \citet{2021MNRAS.506.2603M}.
We divide the simulation data into 10 bins in logs according to the halo mass and plot the histogram of the core mass as shown in Fig.~\ref{fig:May_Mc_pdf}. 
The red line is the result of the log-normal fitting, showing that the distribution seems in good agreement with the log-normal function.

\onecolumngrid

\end{document}